\documentclass[12pt]{article}
\usepackage{amsmath,amssymb,amsfonts}
\usepackage[dvips]{graphicx}
\usepackage{epsfig}
\usepackage{sectsty}
\usepackage{cite}

\makeatletter \@addtoreset{equation}{section} \makeatother

\allsectionsfont{\large\bfseries}
\renewcommand{\thefootnote}{\#\arabic{footnote}}
\def\IC{\mathbb{C}}
\def\IP{\mathbb{P}}
\def\IR{\mathbb{R}}\def\IZ{\mathbb{Z}}

\def\CL{{\cal L}}
\def\CO{{\cal O}}

\def\a{\alpha}\def\b{\beta}\def\g{\gamma}
\def\d{\delta}\def\e{\epsilon}
\def\th{\theta}
\def\l{\lambda}
\def\m{\mu}\def\n{\nu}
\def\r{\rho}\def\s{\sigma}

\def\O{\Omega}

\def\half{\frac{1}{2}}

\def\goto{\rightarrow}

\def\p{\partial}
\def\tr{{\rm tr}}

\newcommand{\ket}[1]{\vert  #1\rangle}
\newcommand{\bra}[1]{\langle #1 \vert}
\newcommand{\vev}[1]{\left\langle{#1}\right\rangle}

\newcommand{\be}{\begin{eqnarray}}
\newcommand{\ee}{\end{eqnarray}}
\newcommand{\nn}{\nonumber}
\newcommand{\bn}{\begin{enumerate}}
\newcommand{\en}{\end{enumerate}}

\begin{document}

\begin{titlepage}
\vfill
\begin{flushright}
{\tt\normalsize KIAS-P08062}\\
{\tt\normalsize SU-ITP-08-21}\\
\end{flushright}
\vfill
\begin{center}
{\large\bf A Nonperturbative Test of  M2-Brane Theory}

\vfill
Kazuo Hosomichi$^{1}$\, Ki-Myeong Lee$^{1}$, Sangmin Lee$^{2}$,\\
Sungjay Lee$^1$, Jaemo Park$^3$  and Piljin Yi$^{1}$

{\small\it
\vskip 5mm
 $^1$Korea Institute for Advanced Study, Seoul 130-722, Korea \\
 $^2$Dept.\!\! of Physics \& Astronomy, Seoul National University,
     Seoul 151-747, Korea \\
 $^3$Dept. of Physics and PCTP, POSTECH, Pohang 790-784, Korea \\
 $^3$Dept. of Physics, Stanford University, Stanford, CA 94305-4060, USA

}
\vfill
\end{center}

\begin{abstract}
\noindent
We discuss non-perturbative effects in the ABJM model
due to monopole instantons. We begin by constructing the
instanton solutions in the $U(2)\times U(2)$ model, explicitly,
and computing the Euclidean action. The Wick-rotated
Lagrangian is complex and its BPS monopole instantons are
found to be a delicate version of the usual 't Hooft-Polyakov
monopole solutions. They are generically 1/3 BPS but become
1/2 BPS at special locus in the moduli space of two M2-branes, yet
each instanton carries eight fermionic zero modes, regardless of
the vacuum choice. The low energy effective action
induced by monopole instantons are quartic order in derivatives.
 The resulting vertices
are nonperturbative in $1/k$, as expected, but are rational
functions of the vacuum moduli. We also analyze the system of
two M2-branes in the supergravity framework and compute the
higher order interactions via 11-dimensional supergraviton exchange.
The comparison of the two shows that the instanton vertices
are precisely reproduced by this M2-brane picture, supporting
the proposal that the ABJM model describes multiple M2-branes.

\end{abstract}

\vfill
\end{titlepage}

\renewcommand{\thefootnote}{\#\arabic{footnote}}
\setcounter{footnote}{0}

\section{Introduction and Summary }

Understanding the worldvolume dynamics of M2-branes is an
important step in the study of M-theory. As a particularly
interesting application, the superconformal field theory on the
worldvolume of multiple M2-branes is believed to give a
holographic description of the eleven-dimensional quantum
supergravity on $AdS_4\times S^7$. A supergravity analysis showed
\cite{Klebanov:1996un} that the number of degrees of freedom on
$N$ M2-branes scales like $N^{3/2}$, which implies a nontrivial
interaction between the coincident M2-branes. This peculiar
scaling property was believed to show up in the infrared strong
coupling limit of the super Yang-Mills theory on $N$ D2-branes,
although so far we have been unable to get a precise understanding
of its origin from the microscopic viewpoint.

It has  been realized that the Chern-Simons gauge theories can
have higher supersymmetries than the familiar ${\cal N}=3$ barrier
once the Yang-Mills term is turned off, and the resulting
Chern-Simons-matter theories may have applications to multiple
M2-branes. Especially, a  maximally supersymmetric Chern-Simons
matter theory  has been constructed by Bagger, Lambert
\cite{Bagger:2006sk,Bagger:2007jr,Bagger:2007vi} and Gustavsson
\cite{Gustavsson:2007vu,Gustavsson:2008dy} based on a
mathematical structure called 3-algebra.
On the other hand, another series of works
\cite{Gaiotto:2008sd,Hosomichi:2008jb,Aharony:2008ug,Hosomichi:2008jd}
based on a more conventional approach have led to the full
classification of Chern-Simons matter theories with
${\cal N}=4,5,6$ supersymmetry. See also
\cite{Bagger:2008se,Bandres:2008ry,Schnabl:2008wj,Bergshoeff:2008bh}.

A particularly interesting example, called the ABJM model
\cite{Aharony:2008ug}, is an ${\cal N}=6$ superconformal
Chern-Simons matter theory where the $U(N)\times U(N)$ gauge
fields of Chern-Simons level $(k,-k)$ are coupled to
bi-fundamental matters. Aharony et.al. \cite{Aharony:2008ug}
proposed that this model is the worldvolume theory of $N$
M2-branes in the orbifold ${\mathbb C^4}/{\mathbb Z}_k$. There are
a number of evidences supporting this proposal from the analysis
of vacuum moduli space, brane construction, etc. Further analysis
has been made on its mass deformation
\cite{Gomis:2008cv,Hosomichi:2008qk,Hosomichi:2008jb,Gomis:2008vc}
and the effect of fractional M2-branes \cite{Aharony:2008gk}.
Recently, a perfect agreement of the superconformal index between
the field theory and the dual supergravity was found in a certain
limit \cite{Bhattacharya:2008bja}, and further evidences
supporting the proposal have been found in the integrability
structure of the two theories
\cite{Nishioka:2008gz,Gaiotto:2008cg,Grignani:2008is,Minahan:2008hf,Arutyunov:2008if,
Stefanski:2008ik,Gromov:2008bz,Gromov:2008qe,Ahn:2008aa,Bak:2008cp}.

In this paper we make a first step to understand the quantum
correction in the ABJM model at the nonperturbative level, as a
rather nontrivial test of the proposal. In particular
we consider instanton processes in the field theory side and identify
their counterpart in the dual
11-dimensional supergravity approach. To summarize the result first,
we find that instantons in the  ABJM theories are of monopole type
with eight fermionic zero modes each, and that the instanton processes
generate a series of higher order interaction terms in the Coulomb phase.
These range from a four-derivative bosonic terms to eight fermion
vertices. We also find that these higher order correction terms
have a well-understood origin in terms of M2-branes interacting
via supergravity and thereby compute the bulk counterpart accurately.
Finally we show that the scaling behavior of the latter matches
precisely the effective and nonperturbative Lagrangian we computed
from the monopole instanton, which suggests strongly that
this ABJM theory is indeed the worlvolume theory of multiple M2-branes.

There hasve been similar considerations for three-dimensional
${\cal N}=8$ Yang-Mills theory\cite{Polchinski:1997pz}.
Here the monopole instanton corrections were interpreted in the
supergravity side as exchanges of D0-branes between a pair
of M2-branes transverse to the M-theory circle, or equivalently
between a pair of D2-branes. Structures of the
resulting higher order corrections were determined quite
precisely \cite{Paban:1998mp,Hyun:1998qf,Hyun:1999hm},
and the match between the Yang-Mills side and the
M-theory side were demonstrated  convincingly.

There are some notable differences between these two cases.
{}From the gravity side, the main difference is in the
eleven-dimensional backgrounds. The former has two M2-branes
in ${\mathbb R}^8/{\mathbb Z}_k\times {\mathbb R}^{2+1}$, while the latter has two M2-branes
(transverse to $S^1$) in $S^1\times {\mathbb R}^7\times {\mathbb R^{2+1}}$.
The D0-branes, which are the bulk counterpart of the
Yang-Mills monopole instantons in the latter, must be now
reinterpreted in the orbifold case, given the absence of a topological
circle, as one of the angular momentum in ${\mathbb R}^8$. The
angular momentum in question turned out to be along
the direction of the orbifolding action ${\mathbb Z}_k$.

In the field theory side, the difference runs much deeper.
Since the ABJM theory contains a pair of Chern-Simons terms,
 one generally expects a rather different behavior
of monopole instantons, if there is any. For instance,
the Wick-rotated Lagrangian for such theories is not real
 since the Chern-Simons term acquire a factor of $i$.
In part due to this, one generically finds that some real
fields take complex configurations for the saddle point.
However, this is not really a  problem as long as the solution
is regular and converges to the (real) vacuum asymptotically.
Using a complex saddle point here is no different than using
a complex saddle point when we perform ordinary contour
integral of a function with critical points off the real axis.
As long as we make sure the semi-classical configuration
approaches the correct (real) vacuum and as long as we take care
not to over-count excitations around this Euclidean solution,
this is a right thing to do.

Another, potentially more serious, worry arises from the gauge
variance of the classical action.  In monopole backgrounds of
any Chern-Simons theories, asymptotically nontrivial
gauge transformations  shift the Euclidean action by
some imaginary constants. As was argued in \cite{Affleck:1989qf},
naive integration over this gauge orbit
seems to project out the amplitudes involving nonzero number of
monopole-instantons. We show, however, that this argument is
misleading. The gauge variance of the action simply means that the
monopole-instanton carries the unbroken gauge charge, and that it
mediates transitions between states with different
charges\cite{Lee:1991ge}. Gauge variance of monopole action
cancels against the gauge variance due to the two mismatching electric
charges in the initial and the final wave functions, so that
the transition amplitudes are gauge invariant as a whole.

Those who are familiar with Chern-Simons theories may wonder
whether there is a finite action  monopole instanton at all,
since, for example, generic Chern-Simons Yang-Mills theories
are massive gauge theory and cannot have finite action monopole
instantons. There is a well-known linear divergence. If such a
behavior were found here, this by itself would have ruled out
the ABJM model as a theory of M2-branes. Fortunately, however,
there is no such divergence here. In fact, the monopole
instanton solutions here are essentially the
usual BPS monopoles of Yang-Mills theory up to a complexified
gauge rotation. See Section 3 and 5 for the explicit forms.
Our `complexified' monopole instantons are novel and original.
Their nonperturbative effect remains to be explored in less
supersymmetric varieties of the ABJM  type theories.

Our M-theory dual calculation leads to a very detailed and
precise effective Lagrangian for the M2-branes and contains
both perturbative and nonperturbative corrections when viewed
from the field theory side. We have reproduced the correct
scaling behaviors of those corresponding to the nonperturbative
parts by studying monopole instantons, but
stopped short of computing loop corrections to these
interaction vertices, such as loop correction to the
monopole instanton saddle point. Nor did we try to evaluate
the simple perturbative loop corrections, which according
to the M-theory computation, should also begin at the
four-derivative level. It would be interesting to
reproduce the entire structure and the coefficients of M-theory
result, from a purely field theoretic calculation of the
ABJM model.

In Section 2, we start with a brief review of the ABJM model,
focusing especially on its vacuum moduli space.
We present generic vacua of the theory with gauge group
$U(2)\times U(2)$ and their massive spectrum.
Then we turn to study the monopole equations in
$U(2)\times U(2)$ model.
We are able to find the `BPS monopole
instanton configuration' throughout the vacuum moduli space by
simple embedding of the well known 't Hooft Polyakov solution.
Our stationary monopole solution is generically complex.
This monopole solution is 1/3 BPS
at generic points in moduli space, while at some special locus it
becomes 1/2 BPS and takes a simpler form.
In Section 3 we first give the construction of the 1/2 BPS solutions
in the special cases where only one complex
scalar field takes nonzero vacuum expectation value,
and then present the general solutions in Section 5.

In Section 4 we get a simple expression for the Euclidean action,
and find that each monopole-instanton carries eight fermion
zero-modes. Although the action is not invariant under a certain
gauge transformation, it simply reflects the gauge charge carried
by the monopole and does not mean their effects are projected out.
In Section 5, we also calculate the monopole action and the
zero-modes for the general 1/3 BPS multi-monopole instantons.

The issues on gauge invariance in Chern-Simons theories will be
explained in greater detail in Section 6.
Later in that section we also argue that the
instanton effects are described by local vertex operators
in the low-energy moduli dynamics, and discuss
several constraints on their possible forms.

Finally, in Section 7 we move to the M/string theory framework.
We first study the system of two M2-branes in supergravity
and see the correspondence between certain transverse momentum
exchanges between the M2-branes and
the multi-monopole instanton processes in the  ABJM model.
We then turn to the type IIA picture and show that the D0-brane
exchange along the Euclidean geodesic line between two
D2-branes reproduce the monopole instanton action of the field
theory. We also get the correct mass spectrum
in the generic vacua of the field theory from the energy of the
fundamental string connecting two D2-branes.
These agreements between the ABJM model and the dual supergravity
provide strong evidences that the ABJM model is the correct
theory of M2-branes on the orbifold $\mathbb{ C}^4/\mathbb{Z}_k$.

In appendix A we take the simple example of abelian BF-matter theory,
and give the explicit construction of the so-called 0-cocycle
which is necessary to make the action gauge invariant.
This is complementary to the abstract discussion of cocycles
given in Section 6.
In Appendix B, we
recalculate the Euclidean action of
monopole instantons by somewhat different approach
from Section 4. In Appendix C, we recapitulate the monopole vertex
operators in the Maxwell theory and in the Chern-Simons matter theory
for further clarification.

\section{The ABJM Model}

We present in this section a short description on the ABJM model
\cite{Aharony:2008ug}, believed to describe the dynamics of
multiple M2-branes probing a certain orbifold geometry. This
${\cal N}=6$ supersymmetric model has the gauge symmetry $G=
U(N)_1\times U(N)_2$ whose gauge fields are denoted by $A_\mu$ and
$\tilde A_\mu$ with the Chern-Simons kinetic term of level
$(k,-k)$. The matter fields are composed of four complex scalars
$Z_\a$ ($\a=1,2,3,4$) and four three-dimensional spinors
$\Psi^\a$, both of which transform under $G$ as $({\bf N}, \bar
{\bf N})$. As well as the gauge symmetry, the present model also
has additional global  $SU(4)$ R-symmetry, under which the scalars
$Z_\alpha$ furnish the representation ${\bf 4}$ while the
fermions $\Psi^\alpha$ furnish $\bar {\bf 4}$.

Let us start with the Lagrangian of the ABJM model,
\begin{eqnarray}
  {\cal L}={\cal L}_\text{CS} + {\cal L}_\text{kin} +
  {\cal L}_\text{Yukawa} + {\cal L}_\text{potential} \ ,
\end{eqnarray}
where
\begin{eqnarray}
  {\cal L}_\text{CS} + {\cal L}_\text{kin}
  &=& \frac{k}{4\pi}\e^{\mu\nu\rho}
  \text{tr} \big( A_\mu \partial_\nu A_\rho - i \frac{2}{3}
  A_\mu A_\nu A_\rho  - \tilde{A}_\mu \partial_\nu \tilde{A}_\rho
  + i \frac{2}{3}  \tilde{A}_\mu \tilde{A}_\nu \tilde{A}_\rho \big)
  \nonumber \\ &&
  \hspace{0.3cm}- \text{tr}  \left( D_\mu \bar{Z}^\a D^\mu Z_\a -
  i \bar{\Psi}_\a \g^\mu D_\mu \Psi^\a \right) \ ,  \nonumber \\
  {\cal L}_\text{Yukawa}  &=&
  -  \frac{2\pi i}{k}\text{tr}\left( \bar{Z}^\a Z_\a \bar{\Psi}_\b \Psi^\b
  - Z_\a \bar{Z}^\a \Psi^\b \bar{\Psi}_\b
  +2 \bar{Z}^\a \Psi^\b \bar{\Psi}_\a Z_\b
  -2 Z_\a \bar{\Psi}_\b \Psi^\a \bar{Z}^\b \right) \nonumber \\
  && \hspace{0.3cm} - \frac{2\pi i}{k} \epsilon^{\a\b\g\d} \text{tr}
  \left(Z_\a \bar{\Psi}_\b Z_\g \bar{\Psi}_\d \right)
  +  \frac{2\pi i}{k} \epsilon_{\a\b\g\d} \text{tr}
  \left(\bar{Z}^\a \Psi^\b \bar{Z}^\g \Psi^d \right)\ ,
\end{eqnarray}
and
\begin{eqnarray}\label{spotential}
  {\cal L}_\text{potential} && \hspace{-0.5cm} = +
    \frac{4\pi^2}{3k^2} \text{tr}
  \Big( Z_\a \bar{Z}^\a Z_\b \bar{Z}^\b Z_\g \bar{Z}^\g +
  \bar{Z}^\a Z_\a \bar{Z}^\b Z_\b \bar{Z}^\g Z_\g \nonumber \\
  && \hspace{1.5cm} + 4 Z_\a \bar{Z}^\g Z_\b \bar{Z}^\a Z_\g \bar{Z}^\b
  - 6 Z_\a \bar{Z}^\a Z_\b \bar{Z}^\g Z_\g \bar{Z}^\b \Big)\ .
\end{eqnarray}
We basically use the convention of \cite{Hosomichi:2008jd}
except the hermitian gauge fields so that the covariant
derivatives now become
\begin{eqnarray}
  D_\mu Z_\a = \partial_\mu Z_\a - i A_\mu Z_\a + i Z_\a \tilde A_\mu \ ,
\end{eqnarray}
and Chern-Simons level  $k $ is now quantized as an integer, i.e.,
$k \in \mathbb{Z}$.  The trace is  over $N\times N$ matrices of
either gauge group and leaves the gauge invariant quantities. The
contraction of spinor fields is the standard one. This Lagrangian
is invariant under the ${\cal N}=6$ supersymmetry whose
transformation rules are
\begin{eqnarray}
  \d Z_\a  &=& - i \eta_{\a \b} \Psi^\b,  \\
  \d \Psi^\a  &=& \left[ \g^\mu D_\mu Z_\g - \frac{4\pi}{3k}
  (Z_{[\b} \bar{Z}^\b Z_{\g]} )  \right]\eta^{\g\a}
  + \frac{8\pi}{3k} (Z_{\b} \bar{Z}^\a Z_{\g}) \eta^{\g\b}
  - \frac{4\pi}{3k} \e^{\a\b\g\d} (Z_{\b} \bar{Z}^\r Z_{\g}) \eta_{\d\r}\ ,
  \nonumber \\
  \d A_\mu &=&  \frac{2\pi i}{k} ( \eta^{\a\b} \g_\mu Z_\a \bar{\Psi}_\b
  + \eta_{\a\b} \g_\mu \Psi^\b \bar{Z}^\a ), \ \
  \d \tilde A_\mu  ~=~
 \frac{2\pi i}{k} (\eta^{\a\b} \g_\mu \bar{\Psi}_\b Z_\a
  + \eta_{\a\b} \g_\mu \bar{Z}^\a \Psi^\b )\ , \nonumber
  \label{susytrans} \end{eqnarray}
where the transformation parameters $\eta^{\a\b}$ satisfy the relations
\begin{eqnarray}
  \eta^{\a\b} = - \eta^{\b\a}, \qquad
\eta_{\a\b} =(\eta^{\alpha\beta})^* = \half \e_{\a\b\g\d} \eta^{\g\d}\  .
\label{12susy} \end{eqnarray}

Let us now examine the vacuum moduli space of the present model at
the classical level, i.e., solutions of $V(\Phi)=0$ up to gauge
transformations. It is known that the potential can be made into a
sum of squares
\begin{eqnarray}
  V =   \frac{2\pi^2}{3k^2} \text{tr} \Big(
  W_{\alpha\gamma}^{~\beta} \bar W^{\gamma\alpha}_{~\beta}
  \Big)
\end{eqnarray}
with
\begin{eqnarray}
  && W_{\alpha\gamma}^{~\beta} = (2Z_\alpha\bar Z^\beta Z_\gamma
  -\delta^\beta_\gamma Z_\alpha\bar Z^\rho Z_\rho
  -\delta^\beta_\alpha Z_\rho  \bar Z^\rho Z_\gamma)
  - (\a \ \leftrightarrow \  \g)\ , \nonumber \\
  && \bar W^{\alpha\gamma}_{~\beta} = (2\bar Z^\alpha Z_\beta\bar Z^\gamma
  -\delta_\beta^\gamma\bar Z^\alpha Z_\rho\bar Z^\rho
  -\delta_\beta^\alpha\bar Z^\rho   Z_\rho\bar Z^\gamma)
  - (\a \ \leftrightarrow \  \g)\ ,
\end{eqnarray}
which leads to the equation for its minima
\begin{equation}\label{vac}
 Z_\alpha \bar Z^\beta Z_\gamma = Z_\gamma \bar Z^\beta Z_\alpha\,,~~~
 \bar Z^\alpha Z_\beta \bar Z^\gamma = \bar Z^\gamma Z_\beta \bar Z^\alpha\,.
\end{equation}
This implies that the hermitian matrices $Z_\alpha\bar Z^\beta$
commute with each other, and similarly for $\bar Z^\alpha Z_\beta$.
The vacuum solutions are thus
given by diagonal $Z_\alpha$ up to gauge equivalences,
\begin{eqnarray}
  Z_\a  = \text{diag}(z^1_\a, z^2_\a, .. , z^N_\a)\   \label{zdiag} .
\end{eqnarray}
On a generic point of the vacuum moduli space, the gauge group
$G=U(N)\times U(N)$ is spontaneously broken
down to $U(1)^N \subset U(N)_D$, diagonal part of $G$.

In order to describe a classical Lagrangian that governs
the dynamics of massless moduli fields, we first take
the diagonal elements of gauge fields $A_\mu$ and $\tilde A_\mu$, i.e.,
\begin{eqnarray}
  A_\mu = \text{diag}(a_\mu^1, a_\mu^2,..,a_\mu^N),
  \qquad \tilde A_\mu = \text{diag}(\tilde a_\mu^1, \tilde a_\mu^2,..,
  \tilde a_\mu^N)\ .
\end{eqnarray}
Although $a^i-\tilde a^i$ are the gauge fields of the broken gauge
symmetries, we need to keep them~\cite{Lambert:2008et,Distler:2008mk}.
In terms of these diagonal variables, the classical low-energy Lagrangian is
\begin{eqnarray}\label{BF}
  {\cal L}_\text{cl} = - \sum_i |D_\mu z^i_\a|^2 +
  \sum_i \frac{k}{4\pi} \e^{\mu\nu\rho}   (a_\mu^i - \tilde a_\mu^i )
  f^i_{\nu\rho}  \ ,
\end{eqnarray}
where $ D_\mu z^i_\a = \partial_\mu z_\a^i  -i (a^i_\mu - \tilde a^i_\mu )
z^i_\a $ and $f^i=d(a^i + \tilde a^i)/2$.
The role of the Chern-Simons terms for the moduli dynamics can be
seen best by dualizing $ (a_\mu^i + \tilde a_\mu^i)/2$. This is
done by adding to ${\cal L}_\text{cl}$ a term
\begin{eqnarray}
  \CL_\text{dual}= - \frac{1}{4\pi} \e^{\mu\nu\rho}
  \sum_i \partial_\mu \theta^i f^i_{\nu\rho}\ ,
\end{eqnarray}
and by treating $f^i$ as the fundamental variable. The $\theta^i$
variables are normalized to have period $2\pi$. Integrating over
$\theta^i$ brings us back to the original low energy Lagrangian,
whereas integrating over $f^i$ imposes the condition,
\begin{equation}
k(a_\mu^i-\tilde a_\mu^i)=\partial_\mu \theta^i\ .
\end{equation}
The Chern-Simons terms disappear upon this, while the
kinetic term simplifies to an ordinary  linear sigma model
\begin{eqnarray}\label{correctscalar}
  {\cal L} = - \left| \partial_\mu \tilde z_\a^i \right|^2,
  \qquad \tilde z_\a^i = e^{-i\theta^i/k } z_\a^i\ .
\end{eqnarray}
Note that $\tilde z^i_\alpha$ are invariant under local gauge
transformations. The $2\pi$ periodicity of $\theta^i$, combined with
the Weyl symmetry, tells us that the vacuum moduli space is an
orbifold
\begin{eqnarray}
  \left(\mathbb{C}^4/\mathbb{Z}_k \right)^N/S_N\ ,
\end{eqnarray}
and also that the correct low energy variables to use are these
invariant fields $\tilde z^i_\alpha$~\cite{Aharony:2008ug}.
These gauge invariant moduli coordinates $\tilde{z}^i_\alpha,
i=1,\ldots,N$ denote the positions of $N$ M2-branes on the orbifold
$\mathbb{C}^4/\mathbb{Z}_k$ after a proper scaling.

Let us now in turn discuss the vacuum degeneracy of the theory.
Since we can add to a given ground state, without costing any
energy, the magnetic flux $f^i_{12}$ together with certain amounts
of charges, the theory has huge number of vacuum degeneracy. Here
the flux and charges that specify the vacuum should satisfy the
Gauss laws of the model (\ref{BF}),
\begin{eqnarray}
\frac{k}{2\pi}f^i_{12} - i (D_0 z^i_\alpha
\bar{z}^{i\alpha} -z^i_\alpha D_0 \bar{z}^{i\alpha} ) &=& 0\ ,
\nonumber \\
\partial_1 (a^i_2-\tilde a^i_2) -\partial_2 (a^i_1-\tilde a^i_1) &=& 0\ .
\end{eqnarray}
Magnetic monopole instantons are those which interpolate between
vacua of different magnetic flux and charges. The monopole
instantons thus violate some of the global charges in
the vacuum moduli dynamics (\ref{correctscalar}).
In Section 6 we will construct the local vertex operators
describing their effect using the gauge invariant variables
$\tilde z^i_\alpha$.

We close this section with mass spectrum on the generic point of
vacuum moduli space. For an instance, let us consider the vacua of the theory
with $U(2)\times U(2)$ gauge group. By the $SU(4)_{\text{R}}$  and
gauge transformations, one can parameterize them as
\begin{equation}
\label{generic-vev}
\langle Z_1\rangle =  \left(\begin{array}{cc} u_1 & 0 \\ 0&  u_2
\end{array}\right),\qquad
\langle Z_2\rangle=  \left(\begin{array}{cc} cu_2 & 0 \\ 0&  cu_1
\end{array}\right),\qquad  \langle Z_3\rangle = \langle Z_4\rangle = 0\,,
\end{equation}
where  the parameters are all real and obey $0<u_1<u_2$ and $0<c$.
Note that the two M2-brane are at $z^1_\alpha
= (u_1,cu_2,0,0)$ and $z^2_\alpha= (u_2,cu_1,0,0)$. The linear
fluctuation analysis tells us that the mass spectrum in this
vacuum is given by
\begin{eqnarray}
 \text{massless  multiplet} &:& 16 \ \text{scalar bosons}
 + 16 \ \text{fermions} ,  \nonumber \\
 \text{massive multiplet} &:&  12\ \text{scalar bosons}
 +16\  \text{fermions}+ 4 \ \text{vector bosons} ,
 \end{eqnarray}
where the mass of the massive multiplet is
\begin{eqnarray}
\mu  &=& \frac{2\pi}{|k| } \sqrt{\big((z^1 \cdot \bar z^1)^2
+(z^2 \cdot \bar z^2 )^2 \big)^2
- 4 \big| z^1 \cdot \bar z^2 \big|^2} \nonumber \\
&=&
\frac{2\pi}{|k| }(1+c^2)(u_2^2-u_1^2)\ . \label{mass0}
\end{eqnarray}
This agrees with the result in \cite{Berenstein:2008dc}. Here dot
indices denote the $SU(4)_\text{R}$ indices contraction. The spin
structure of the massive multiplet is $(1,\frac12,0,-\frac12,-1)$
with multiplicity $2\times (1,4,6,4,1)$. In Section 7, we
interpret the vacuum expectation value (\ref{generic-vev}) as the
positions of two M2-branes in $\mathbb{C}^4/\mathbb{Z}_k$, and the
M2-brane connecting these two branes has the energy given by the
above mass formula.

\section{Monopole Instantons and the Reality Condition}

In this section, we wish to look for monopole instanton solutions.
For the instanton physics, we  consider the Euclidean version of
the theory. As  usual,  we take the Wick rotation $t=-i\tau$ to
obtain the Euclidean Lagrangian,
\begin{eqnarray}
  - {\cal L}_\text{E} = i {\cal L}_\text{CS} + {\cal L}_\text{kin}
  + {\cal L}_\text{Yukawa} + {\cal L}_\text{potential}\ .
\end{eqnarray}
It is noteworthy here that the Chern-Simons coupling ${\cal
L}_\text{CS}$ gets the imaginary sign which will introduce
several subtle issues in later sections.
Three-dimensional Euclidean gamma matrices
$\g^\mu$ are chosen to satisfy the relations
\begin{eqnarray}
  \left\{ \g^\mu , \g^\nu \right\} = 2 \d^{\mu\nu} ,
  \hspace{0.5cm} \g^{\mu\nu\rho} = i\e^{\mu\nu\rho}\ .
\end{eqnarray}

In the most of this work we focus on the case with
$U(2)\times U(2)$ gauge group, which is the simplest where
monopole instantons appear.
We will work with the parametrization of the vacua given in
(\ref{generic-vev}).
Let us begin with the special case $c=0$ where only one
of the four scalars takes non-zero vev, say $Z=Z_1$,
\begin{equation}
\langle Z\rangle = U \left(\begin{array}{cc} u_1 & 0 \\ 0&  u_2
\end{array}\right)V^{-1}  =\langle \bar Z\rangle^\dagger
\label{vacuum1}
\end{equation}
for some unitary $U$ and $V$. $\langle\bar Z\rangle$ is of course
the conjugate of $\langle Z\rangle$, so the latter equation is
redundant. The reason we show it explicitly should become clear in
a moment. Without loss of generality, we suppose  that $u_{1,2}$
are real and that $0<u_1<u_2$. In terms of the M2-brane
interpretation, these two are radial positions of the two M2-branes
in the orbifold $\mathbb{C}^4/\mathbb{Z}_k$.

We are looking for a monopole instanton that preserves some
supersymmetry. The BPS equation coming from supersymmetry
transformation is pretty simple when we turn on only one
of the four scalar fields, and with
\begin{equation}
D  Z\equiv  d  Z-iA  Z+iZ\tilde A  \quad \hbox{and}
\quad D  \bar Z\equiv d   \bar Z-i\tilde A   \bar Z+i\bar ZA\ ,
\end{equation}
we have
\begin{equation}
D  Z=0\quad\hbox{\it or} \quad D  \bar Z=0
\label{halfBPS}
\end{equation}
as the condition for half-BPS configurations.
One would think that the second equation
is the same as the first, again since $\bar Z$ is merely a
conjugate of $Z$, in which case this will certainly lead to
constant $Z$ and $\bar Z$ only.

However, the ABJM model is a Chern-Simons theory. The
Chern-Simons term acquires a factor $i$ upon Wick rotation, and
the Euclidean action becomes complex. In such circumstances, the saddle
point evaluation can often involve deformation of the
path-integral into complex planes of (what used to be real) field
variables. In Appendix B, such complexified stationary path is found for
a very simple mechanical model.
The semi-classical configurations that dominate the
path integral need not satisfy the usual reality constraints. This
is nothing new, and we do such deformation of contour all the time
when we perform ordinary integration of real functions.

It may happen that there exists a saddle point where only one of the two
conditions (\ref{halfBPS}) is satisfied, say
\begin{equation}
D  Z=0\ .
\end{equation}
This is the type of saddle points we are interested in, and the
solution we obtain can be interpreted as a monopole-like
instanton.\footnote{ We can treat $D\bar Z $ and $DZ$ differently,
in part because each enters the supersymmetry transformation rule
of $\Psi$ and $\bar\Psi$. In Euclidean signature, as is well known,
these two fermions must be treated as independent variables, so
their supersymmetry transformation can be treated independently as
well.} The broken supersymmetry generators are along
$\eta_{12},\eta_{13},\eta_{14}$ for this
case.\footnote{
Here we assume that the Euclidean supersymmetry parameters
satisfy the reality condition similar to (\ref{12susy}),
implying twelve real supersymmetries in the Euclidean theory as well.}
As it will become clear soon, the other choice $D \bar Z=0$ with broken
supersymmetry generators along $\eta^{1\alpha}$
corresponds to anti-monopole solution.

Using $D Z=0$ together with the Gauss constraints for $A$ and $\tilde A$,
we find the following set of equations
\begin{eqnarray}\label{master}
\frac{k}{2\pi}*F\equiv\frac{k}{2\pi}*(dA-iA\wedge A)&=&- D(Z\bar Z)\ ,
 \nonumber\\
\frac{k}{2\pi}*\tilde F \equiv\frac{k}{2\pi}*(d\tilde A-i \tilde A
 \wedge \tilde A)&=& - D(\bar Z Z)\ .
\end{eqnarray}
Note that
\begin{equation}
 D\! * \! D (Z\bar Z)=0=D\! *\! D (\bar Z Z)
\end{equation}
follows by a further use of the Bianchi identity, so the BPS
equation together with the Gauss constraint implies the equation of
motion
\begin{equation}
D\! * \! D  \bar Z=0
\end{equation}
as long as the covariantly constant $Z$ is nonsingular.

The master equations (\ref{master}) look like ordinary BPS
equation for  monopoles. As an initial attempt,
let us consider $A=\tilde A $, so that $F=\tilde F
$. The BPS equation then implies $ D\wedge D  Z=-i[F,
Z]=0$, which together with (\ref{master}) forces
(with some constants $a,b,c$)
\begin{equation}
Z=c {\bf 1}_2, \qquad \bar Z=a\,\Phi+b {\bf 1}_2 ,
\end{equation}
where $\Phi$ is a $2\times 2$ traceless scalar function that,
together with $A=\tilde A$, solves the ordinary monopole BPS
equation. However, this has the asymptotic value $\langle
Z\rangle^\dagger \neq \langle \bar Z\rangle$ which violates the
reality condition, and, as such, is unusual.  The only exception
occurs when $a=0,b^*=c$ which brings us back to a vacuum.

Underlying this difficulty is that the gauge fields $A=\tilde A$
in this ansatz is perfectly real, even though we do not expect the
saddle point that obeys usual reality conditions. What we cannot
abandon is the reality condition of the vacuum itself, so we must
be prepared to trade off the (partial) reality of the instanton
solution in favor of the reality of the scalar vev.

Motivated by this initial failure, let us consider the
following redefinition of variables
\begin{equation}
Z=L{\cal Z}L,\quad \bar Z=L^{-1}\bar{\cal Z}L^{-1}
\end{equation}
accompanied by cancelling transformation of the gauge fields,
\begin{equation}
A=L{\cal A}L^{-1}+iLdL^{-1}\;,\quad
\tilde A=L^{-1}{\cal \tilde A}L+iL^{-1}dL\;,
\end{equation}
none of which preserve the reality conditions. On the other hand,
the BPS equation and the Gauss constraint are preserved, so
${\cal A, \tilde A, Z,\bar Z}$ obey the
same set of equations as ${  A, \tilde A, Z,\bar Z}$. One can think
of $L$ as a complexified
gauge transformation, although we are not suggesting it as a symmetry
of the theory itself.

The point of doing this redefinition is that now we can
use the ansatz ${\cal A}=\tilde {\cal A}$ without worrying about the
reality condition between $\langle{\cal Z}\rangle$ and $\langle
\bar {\cal Z}\rangle$.
The general solution with the reality condition
$\langle Z\rangle = \langle \bar Z\rangle^\dagger$
asymptotically satisfied turns out to be
\begin{equation}\label{scalars0}
{\cal Z}= \sqrt{u_1 u_2} {\bf 1}_2\ , \qquad  \bar {\cal Z}
= \frac{1}{\sqrt{u_1 u_2}} \left(  (u_1^2 - u_2^2)\,\Phi
+ \frac{u_1^2 + u_2^2 }{2} {\bf 1}_2 \right),
\end{equation}
and
\begin{eqnarray}\label{bps0}
*{\cal F}\equiv*(d{\cal A}-i{\cal A}\wedge {\cal A})&=&
\mu\left(d\Phi-i[{\cal A},\Phi]\right)\,,
\end{eqnarray}
where $\Phi$ is normalized so that $\tr\langle \Phi\rangle^2=1/2$
and  $\mu$ is the mass
  parameter (\ref{mass0}) with $c=0$,
\begin{eqnarray}
  \mu = \frac{2\pi}{k} \left( u_2^2 - u_1^2 \right) >0.
\end{eqnarray}
The equation (\ref{bps0}) is nothing but the usual BPS
monopole equation with the scale $\mu$ \cite{Weinberg:2006rq}.
The solution for a single monopole is
\begin{eqnarray}
  \Phi = \left( \coth{\mu r} - \frac{1}{\mu r} \right)
  \frac{\hat r^a   \sigma^a}{2},
  \hspace{0.5cm} \tilde{\cal A}={\cal A}= \frac12
  \left(\frac{\mu r}{\sinh{\mu r}}-1 \right) \e^{abc}\s^a \hat r^b d \hat r^c\ .
\end{eqnarray}
One can reconstruct $A,\tilde A, Z,\tilde Z$ by finding
appropriate transformation matrix $L$.

To  find $L$, and also to see how (\ref{scalars0}) leads to the solution
with physically acceptable vev, consider
\begin{equation}\label{ansatz}
  Z  = \sqrt{u_1 u_2}L^2  ,\quad  \bar Z  =
L^{-1} \frac{1}{\sqrt{u_1 u_2}} \left(  (u_1^2 - u_2^2)\,
 \Phi  + \frac{u_1^2 + u_2^2 }{2} {\bf 1}_2 \right) L^{-1} \ .
\end{equation}
With real $u_{1,2}$ it is clear that
$\langle Z\rangle =\langle\bar Z\rangle^\dagger$
can be satisfied for $L$ of the general form
\begin{equation}
L=e^{\Lambda(x)  \langle\Phi\rangle }\ , \label{leq0}
\end{equation}
where asymptotic value $\Lambda_*$ of $\Lambda(x)$
is constant on    $S_\infty^2$. This value  should be
\begin{equation}
e^{\Lambda_*}=\sqrt{\frac{u_1}{u_2}} . \label{leq1}
\end{equation}
To see this,  we need to compare the asymptotic value
at each point on $S^2_\infty$.
This can be easily done in the unitary gauge
$\langle \Phi\rangle=\sigma_3/2$ where we have
 \begin{equation}
\lim_{x\rightarrow\infty}L^2=e^{\Lambda_* \sigma_3}\ ,   \end{equation}
and (\ref{ansatz}) leads to the vev
\begin{equation}
\langle Z\rangle =
\left(\begin{array}{cc} u_1 & 0 \\ 0&  u_2 \end{array}\right)
=\langle \bar Z\rangle^\dagger
\end{equation}
as promised, up to gauge rotations $U$ and $V$.
One choice of $L$ which is smooth everywhere
is  $ L = e^{ -\frac12\log( u_2/u_1)\, \Phi(x) }$.

Finally, let us consider the other choice of BPS equation $D\bar Z=0$.
This choice leads to a different set of equations when combined with
the Gauss constraints
\begin{eqnarray}\label{master1}
\frac{k}{2\pi}*F\equiv\frac{k}{2\pi}*(dA-iA\wedge A)&=& D(Z\bar Z)\ ,
\nonumber\\
\frac{k}{2\pi}*\tilde F \equiv\frac{k}{2\pi}*(d\tilde A-i \tilde A
\wedge \tilde A)&=&  D(\bar Z Z)\ .
\end{eqnarray}
The analog of (\ref{ansatz}) for the scalar field is now
\begin{equation}\label{scalars}
 Z=\tilde L \frac{1}{\sqrt{u_1 u_2}} \left(  (u_1^2 - u_2^2)\,\Phi
+ \frac{u_1^2 + u_2^2 }{2} {\bf 1}_2 \right)\tilde L,\qquad
 \bar Z
=\sqrt{u_1 u_2}\;\tilde L^{-2},
\end{equation}
with $A=\tilde L{\cal A}\tilde L^{-1}+i\tilde Ld\tilde L^{-1},\;
\tilde A=\tilde L^{-1}{\cal \tilde A}\tilde L+i\tilde L^{-1}d\tilde L$,
which leads us to the anti-BPS equation for ordinary monopoles
\begin{eqnarray}\label{bps}
*{\cal F}\equiv*(d{\cal A}-i{\cal A}\wedge {\cal A})&=&
-\mu\left(d\Phi-i[{\cal A},\Phi]\right)
\end{eqnarray}
with the same scale $\mu>0$ as before. If we choose to write the
anti-monopole instanton to have the same $\langle\Phi\rangle$ as that of
the monopole instanton, $\tilde L=L^{-1}$ will do the trick for
reconstruction of the anti-monopole instanton $A,\tilde A,Z,\bar Z$
from this data. What is important for us is that the two cases
differ by $Z \leftrightarrow \bar Z$ and the relative sign change
between $d{\cal A}-i{\cal A}^2$ and $(d-i{\cal A})\Phi$.

\section{Euclidean Action and Zero-Modes}

There is a potential subtlety with the Euclidean action, because
a Chern-Simons monopole mediates two states that differ by $k$ units
of electric charge. When the transition is not vacuum-to-vacuum one,
the computation of the WKB amplitude can in general involve the
so-called cocycle factor. However, at the end of the day the ordinary
Euclidean action would suffice with the present solution, as far as the
modulus of the WKB amplitude goes, so let us evaluate $S_{\rm E}$ for our
solutions. The cocycle issues will be addressed in Section 6
and Appendix A.
An alternative evaluation of the monopole action is given in Appendix B.

$S_{\rm E}$ has three bosonic pieces, the Chern-Simons term, the
scalar kinetic term, and the potential term. The potential term
does not contribute since only one complex scalar is turned on, while
the scalar kinetic term, $D\bar Z^\a DZ_\a$, vanishes on either of BPS
or anti-BPS equations, $DZ=0$ or $D\bar Z=0$. Thus, the only piece that
contributes is the Euclidean Chern-Simons action. For a monopole
instanton, therefore, we find
\begin{eqnarray}
-S_{\rm E}&=&\frac{ik}{4\pi}\int\left(\omega_3(A)-\omega_3(\tilde A)\right)
 \nonumber\\
&=&\frac{ik}{4\pi}\int \left(\omega_3(L{\cal A}L^{-1}+iLdL^{-1})
-\omega_3(L^{-1}\tilde{\cal  A}L+iL^{-1}dL)\right).
\label{S_E}
\end{eqnarray}
This can be split into pieces involving ${\cal A =\tilde A}$ only,
which cancel each other, and the rest
\begin{equation}
-S_{\rm E}=\frac{k}{4\pi}\int_{S^2_\infty}
 \left(\int_0^1 ds\,\tr [\log(L)dA_s]\right)
+\frac{k}{4\pi}\int_{S^2_\infty} \left(\int_0^1 ds\,
 \tr [\log(L)d\tilde A_s]\right)
\end{equation}
with $A_s\equiv L^s{\cal A}L^{-s}+iL^sdL^{-s}$ and
$\tilde A_s\equiv L^{-s}{\cal A}L^{s}+iL^{-s}dL^{s}$.
Thus, it suffices to understand the asymptotic behavior of the gauge fields.

Parameterizing $\langle \Phi\rangle$ as $n^a\sigma^a/2$ with a unit
3-vector $n$, the asymptotic gauge field has the form,
\begin{equation}
 {\cal A}\Big\vert_{S^2_\infty}=\frac{\sigma^a}{2}(dn\times n + \alpha n)^a \,,
\end{equation}
where the cross product is with respect to the $SU(2)$ adjoint
indices and $\alpha$ is an arbitrary 1-form. This comes from
$
D\Phi  =O(1/r^2)\,.
$
The asymptotic forms of $dA_s$ and $d\tilde A_s$ are such that
\begin{equation}
n^ad{\cal A}^a\Big\vert_{S^2_\infty}=n^adA_s^a\Big\vert_{S^2_\infty}
=n^ad\tilde A^a_s\Big\vert_{S^2_\infty}= n^a(-dn\times dn)^a+d\alpha
\end{equation}
regardless of $L^s$, since the transformation by $L$ only shifts
$\alpha$ by $\pm id\Lambda$.
It is instructive to consider first
the asymptotic form of $n^a{\cal F}^a$,
\begin{equation}\label{nF}
n^a{\cal F}^a\Big\vert_{S^2_\infty} =\frac12\, n^a(-dn\times dn)^a+d\alpha \,.
\end{equation}
Note the relative
factor $1/2$ in front of the two first terms in the two
expressions. Recall that the monopole solution is such that
\begin{equation}
\int_{S^2_\infty}n^a{\cal F}^a=4\pi
\end{equation}
by definition. For the spherically symmetric Hedge-Hog gauge with
$n^a=-\hat r^a$
and $\alpha=0$, this can be seen explicitly by integrating the first term
of (\ref{nF}).
For more general but still smooth gauge choice, the first term yields
the same $4\pi$ since it is a topological expression while $d\alpha$
should remain exact on $S^2_\infty$. Therefore,
for any smooth gauge choice we find
\begin{equation}
\int_{S^2_\infty} n^ad{\cal A}^a=2\int_{S^2_\infty} n^a{\cal F}^a=8\pi\ .
\end{equation}
The potential subtlety is in the limiting case of the unitary
gauge $n^a=\delta^{a3}$ where $\alpha$ is the Dirac potential of flux
$4\pi$ with a Dirac string. Globally, $d\alpha$ remains exact.
What  happens here is that, in this gauge, the winding number density of
the first term of $n^a{\cal F}^a$ is concentrated along the Dirac
string direction and cancels the Dirac string contribution.
For $n^ad{\cal A}^a$, this does not happen. Instead, the first, winding
term overcompensate the Dirac string piece in $d\alpha$ by a factor of two.
So the Dirac potential (i.e., $d\alpha$ minus the Dirac string) contributes
$4\pi$ and the winding number
density combined with the Dirac string contributes $4\pi$,
so that again we find $\int_{S^2_\infty} n^ad{\cal A}^a=8\pi$.

Therefore, with $L=e^{\Lambda(x)\langle\Phi\rangle}$ and
$\Lambda_*=\Lambda(\infty)$,
the Euclidean action for a single monopole instanton is
\begin{equation}
-S_{\rm E}=2\times \frac{k}{4\pi}\int_{S^2_\infty} \,\Lambda_*\,\tr \left(\langle\Phi\rangle d{\cal A}\right)
=2\times \frac{k\Lambda_*}{8\pi}\int_{S^2_\infty} \,n^a d{\cal A}^a =2k\Lambda_*
\end{equation}
which gives
\begin{equation}
 e^{-S_{\rm E}} ~=~ e^{2k\Lambda_\ast} ~=~ \left(\frac{u_1}{u_2}\right)^k
 ~~~{\rm for}~~~ \Lambda_\ast=\sqrt{\frac{u_1}{u_2}}\;.
\end{equation}
The computation of the Euclidean action for the anti-monopole
instanton proceeds exactly the same manner, except $L$ is replaced
by $L^{-1}$ and ${\cal F=\tilde F}$ has the opposite magnetic
flux. The combined effect is again the same result.
We could have done the same computation for multi-monopole
instantons and multi-anti-monopole instantons, and the result is
\begin{equation}
e^{-S_{\rm E}}=\left(\frac{u_1}{u_2}\right)^{k|m|}
\end{equation}
for the monopole number $m$. Note that our vacuum choice was
such that $0<u_1<u_2$, and the WKB amplitude is suppressed
by powers of $(u_1/u_2)^k$ for each monopole. This is consistent
with $1/k$ as the effective coupling in this theory, for the
amplitude is exponentially suppressed by $k$.
However, the suppression is only powerlike with respect to the
vacuum expectation values.

Now we turn to zero-mode counting.
The number of bosonic zero-modes within the present ansatz
with ${\cal A=\tilde A}$ is clearly $4|m|$ since the problem
collapses to the usual Yang-Mills case. While we do not have
a rigorous proof yet, we believe these usual bosonic zero-modes
of (anti-)BPS monopoles exhaust all such for the monopole instanton
of the present theory. A partial support comes from the fermionic
part of the story, which can be more easily counted.
The fermionic partners, $\Psi^\a$ and $\bar\Psi_\a$ of $Z_\a$
and $\bar Z^\a$, have the following equation of motion when
only $Z=Z_1$ is excited,
\begin{equation}
\frac k{2\pi}
\gamma^\mu D_\mu\Psi^\a \pm (Z\bar Z) \Psi^\a
\mp \Psi^\a (\bar Z Z) =0
\end{equation}
and
\begin{equation}
\frac k{2\pi}
\gamma^\mu D_\mu\bar\Psi_\a \pm
\bar\Psi_\a (Z\bar Z) \mp
 (\bar Z Z) \bar\Psi_\a=0\ ,
\end{equation}
where again, in this Euclidean regime, we treat the
two sets of fermions as independent. The upper sign
is for $\Psi^{2,3,4}$ and $\bar\Psi_{2,3,4}$ while
the lower sign is for $\Psi^{1}$ and $\bar\Psi_{1}$.

Let us  first exploit the general form of monopole
instanton solution, and
go to ${\cal A=\tilde A,Z,\bar Z}$ variables. Redefining
\begin{equation}
\Psi^\alpha=L\psi^\alpha L,\quad
\bar \Psi_\alpha =L^{-1}\bar \psi_\alpha L^{-1}\,,
\end{equation}
the zero-mode equations reduce to
\begin{equation}
\gamma^\mu {\cal D}_\mu\psi^\a \mp \mu [\Phi,\psi^\a]=0
\end{equation}
and
\begin{equation}
\gamma^\mu {\cal D}_\mu\bar\psi_\a \pm \mu [\Phi,\bar\psi_\a]=0\ ,
\end{equation}
where
\begin{equation}
{\cal D}=d-i{\cal A}
\end{equation}
acting on what are effectively the adjoint fermions
$\psi$ and $\bar\psi$. The complication due to the complex
nature of the solution does not enter the index counting
because the scalar contributes only
in terms of  ${\cal Z\bar Z}=-k\mu \Phi/2\pi+(\cdots)\times {\bf 1}_2$.
Note that the constant part ${\cal Z\bar Z}$, proportional
to ${\bf 1}_2$, also disappears since the scalar ${\cal Z\bar Z}$
acts as a commutator.

Thus, the fermion zero-mode problem is reduced to that of
2-component adjoint fermions in ordinary BPS monopole,
$({\cal A}=\tilde{\cal A},\mu\Phi)$,
albeit now in the Euclidean three-dimensional world.
Since the monopole instanton is no longer a solution that
obeys reality condition, the corresponding zero-mode
counting could have been awkward. However, the special form
of the solution $A,\tilde A,Z,\bar Z$ which can be mapped to
${\cal A =\tilde A, Z,\bar Z}$, allows an easy translation to
the zero-mode counting of the ordinary BPS monopole.

The latter says the following:
the field equation for a complex fermion $\psi$ in $m$-monopole background
\begin{equation}
\gamma^\mu{\cal D}_\mu\psi+\mu[\Phi,\psi]=0
\label{psi0}
\end{equation}
has $2m$ zero-modes \cite{Callias:1977kg,Weinberg:1979zt},
whereas the similar equation with the second term
sign-flipped has no zero-modes.
Thus on our one-monopole background we have two zero-modes
from each of $\psi_1$, $\bar\psi_{2,3,4}$.
The transforming matrix $L$ does nothing to the usual normalizability
conditions
on zero-modes, so therefore we have total of eight zero-modes
per each monopole instanton, with two each for
\begin{equation}
\Psi^1,\; \bar{\Psi}_2,\;\bar{\Psi}_3,\;\bar{\Psi}_4\ .
\end{equation}
For anti-monopoles, which also contribute quantum corrections,
the situation is reversed and the roles of $\Psi$ and $\bar \Psi$
are exchanged.

This apparent disparity between $\Psi$ and $\bar \Psi$ is
related to the usual practice of treating them as independent.
What should be remembered, though, is that each zero-mode
of $\Psi$, even though they are complex fields, carries a
single fermionic collective coordinate and likewise for $\bar\Psi$.
Thus, the number of Grassmanian
collective coordinates to saturate, in order to have nonvanishing
contribution to the path-integral, is eight. The vertex
operators one can compute directly from the dilute gas approximation
of monopoles and anti-monopoles should have eight fermions, of
the form
\begin{equation}
(\Psi^1)^2(\bar{\Psi}_2)^2(\bar{\Psi}_3)^2(\bar{\Psi}_4)^2\ .
\end{equation}

\section{General Monopole Instantons, Euclidean Action and Zero-Modes}

So far we considered monopole instanton in a vacuum where only
$Z_1$ takes an expectation value. Even in the simplest of the
ABJM model with $U(2)\times U(2)$, however, this is not the
generic vacuum. As we saw in Section 2, generically three real
parameters  can be turned on, up to the gauge and $SU(4)_{\rm R}$
symmetry transformations, and this forces at least two scalar
fields, say $Z_{1,2}$, take vev as shown in Eq.~(\ref{generic-vev}).
 In such general vacua, the ansatz
we employed above will not work since the general form of the
instanton solution requires turning on at least one more
scalar field, say $Z_2$, in addition to $Z=Z_1$. In particular,
the BPS equation has to be modified to accommodate $Z_2$ and $\bar Z^2$.

The generalized form of the BPS equation with two scalar fields
involved is
\begin{equation}
DZ_1=0\ ,\qquad D\bar Z^2=0\ .
\end{equation}
This preserves one third of the ${\cal N}=6$ supersymmetry with
the preserved supersymmetry parameters  $\eta_{23},\eta_{24}$ of
the supersymmetry transformation (\ref{susytrans}).
Note that a similar choice such as $DZ_1=DZ_2=0$ would lead to the
solutions with $Z_1$ and $Z_2$ proportional to each other,
which are trivially related to the previous 1/2 BPS solutions by a
suitable $SU(4)_{\rm R}$ rotation.
 With this BPS equation, the Gauss constraints reduce to
\begin{eqnarray}
 \frac k{2\pi}\ast F
 \equiv \frac k{2\pi}\ast(dA-iA\wedge A)
 &=& D(Z_2\bar Z^2-Z_1\bar Z^1), \nonumber\\
 \frac k{2\pi}\ast\tilde F
 \equiv \frac k{2\pi}\ast(d\tilde A-i\tilde A\wedge\tilde A)
 &=& D(\bar Z^2Z_2-\bar Z^1Z_1)\,,
\end{eqnarray}
which again suggests a simple mapping to ordinary monopole
BPS equations, except that $\bar Z^2Z_2-\bar Z^1Z_1$ replaces
$-\bar ZZ$.

Recall that we chose the parameterization of the generic vacua
(\ref{generic-vev}) as
\begin{equation}
\label{generic-vev1}
\langle Z_1\rangle =  \left(\begin{array}{cc} u_1 & 0 \\ 0&  u_2
\end{array}\right),\qquad
\langle Z_2\rangle=  \left(\begin{array}{cc} cu_2 & 0 \\ 0&  cu_1
\end{array}\right),\qquad
\end{equation}
where    $0<u_1<u_2$ and $0<c$.
With this, we can again resort to the transformed variables
\begin{eqnarray}
A=L{\cal A}L^{-1}+iLdL^{-1},\quad  \tilde A=L^{-1}{\cal \tilde
A}L+iL^{-1}dL
\end{eqnarray}
and take the ansatz ${\cal A=\tilde A}$. The Gauss constraints
collapse to
\begin{equation}
 \ast {\cal F}\equiv \ast (d{\cal A}-i{\cal A}\wedge{\cal A})
 ~=~ \mu(d\Phi-i[{\cal A},\Phi])\ ,
\end{equation}
where $\mu=\frac{2\pi}k(1+c^2)(u_{2}^2-u_1^2)$ is the mass parameter
for generic vacua (\ref{mass0}).
The monopole scalar function $\Phi$ (with $\tr \langle \Phi^2\rangle =1/2$)
enters the transformed scalar fields as
\begin{eqnarray}
  {\cal Z}_1=\sqrt{u_{1}u_{2}}\,{\bf 1}_2 ,~~
   && \bar{\cal Z}^1=\frac{1}{\sqrt{u_{1}u_{2}}}\left(
    (u_{1}^2-u_{2}^2)\Phi
   +\frac{u_{1}^2+u_{2}^2}2\,{\bf 1}_2
   \right), \nonumber \\
   \bar{\cal Z}^2=c\sqrt{u_1u_2}\,{\bf 1}_2 ,~~
  && {\cal Z}_2=\frac{c}{\sqrt{u_1u_2}}\left(
    (u_{2}^2-u_{1}^2)\Phi
   +\frac{u_{2}^2+u_{1}^2 }2\,{\bf 1}_2
   \right),
\label{sol2Z}
\end{eqnarray}
which is related to the physical scalar fields as,
\begin{equation}
Z_{1,2}=L{\cal Z}_{1,2}L,\quad \bar Z^{1,2}=L^{-1}\bar{\cal Z}^{1,2}L^{-1},
\end{equation}
for some $L$ as before.

An interesting aspect of this solution is that  $L$ is independent of the
constant $c$ and remains
unchanged from that  of the monopole instanton in the
special vacua. Thus $L$ in equations (\ref{leq0}) and (\ref{leq1}) ensures
the reality condition
$\langle Z_{1,2}\rangle =\langle\bar Z^{1,2}\rangle^\dagger$.
Because of this peculiar feature,
which is no doubt due to our nonconventional parameterization
of the vev's,
the Euclidean action of the monopole instanton remains independent
of $c$,
\begin{equation}
\label{SEft}
 e^{-S_{\rm E}} = \left(\frac{u_{1}}{u_{2}}\right)^{k|m|}
\end{equation}
for $m$-monopole instanton in this generic vacuum. We confirm this
$c$-independent action from the M-theory calculation in Section 7.

As suggested by the fact that eight supercharges are broken,
the number of zero-modes remains eight.
Due to $Z_1, Z_2$ being nonzero, the
fermion equation of motion mixes $\Psi^1$ and $\Psi^2$, and also
$\Psi^3$ and $\bar\Psi_4$.  Using that ${\cal Z}_1$ and $\bar{\cal Z}^2$ are
constant and proportional to the identity matrix, we find
that the equations
for $\Psi^3,\Psi^4,\bar\Psi_3,\bar\Psi_4$ become
\begin{eqnarray}
 \frac k{2\pi}\gamma^\mu{\cal D}_\mu\psi^3
 +[{\cal Z}_1\bar{\cal Z}^1+{\cal Z}_2\bar{\cal Z}^2, \psi^3]
 +2[{\cal Z}_1{\cal Z}_2, \bar\psi_4]
 &=& 0, \nonumber \\
 \frac k{2\pi}\gamma^\mu{\cal D}_\mu\psi^4
 +[{\cal Z}_1\bar{\cal Z}^1+{\cal Z}_2\bar{\cal Z}^2, \psi^4]
 -2[{\cal Z}_1{\cal Z}_2, \bar\psi_3]
 &=& 0, \nonumber \\
 \frac k{2\pi}\gamma^\mu{\cal D}_\mu\bar\psi_3
 -[{\cal Z}_1\bar{\cal Z}^1+{\cal Z}_2\bar{\cal Z}^2, \bar\psi_3]
 +2[\bar{\cal Z}^1\bar{\cal Z}^2, \psi^4]
 &=& 0, \nonumber \\
 \frac k{2\pi}\gamma^\mu{\cal D}_\mu\bar\psi^4
 -[{\cal Z}_1\bar{\cal Z}^1+{\cal Z}_2\bar{\cal Z}^2, \bar\psi^4]
 -2[\bar{\cal Z}^1\bar{\cal Z}^2, \psi^3]
 &=& 0
\end{eqnarray}
under $\Psi^\alpha=L\psi^\alpha L$ and
$\bar \Psi_\alpha =L^{-1}\bar \psi_\alpha L^{-1}$.

Recalling ${\cal Z}_1\bar{\cal Z}^1-{\cal Z}_2\bar{\cal Z}^2
 =-k\mu\Phi/2\pi$ up to shifts by a constant multiple of identity
matrix, we find that the following combinations
\[
\psi={\cal Z}_1\bar\psi_3-\bar{\cal Z}^2\psi^4~~~{\rm and}~~~
\psi={\cal Z}_1\bar\psi_4+\bar{\cal Z}^2\psi^3
\]
satisfy the  zero-mode equation (\ref{psi0}). The other
two linear combinations
\begin{eqnarray}
 \psi &=& \sqrt{u_1 u_2}\,
         (u_{1}^2-u_{2}^2)\psi^3
         +c\sqrt{u_{1} u_{2}}
         (u_{2}^2-u_{1}^2)\bar\psi_4\ , \nonumber \\
 \psi &=& \sqrt{u_{1} u_{2}}\,
         (u_{1}^2-u_{2}^2)\psi^4
         -c\sqrt{u_{1} u_{2}}
         (u_{2}^2-u_{1}^2)\bar\psi_3 \nonumber
\end{eqnarray}
satisfy the equation (\ref{psi0}) with the second term
sign-flipped, so that they do not yield zero-modes. The equations
for $\Psi^1,\Psi^2,\bar\Psi_1,\bar\Psi_2$ read
\begin{eqnarray}
 \frac k{2\pi}\gamma^\mu{\cal D}_\mu\psi^1
 -[{\cal Z}_1\bar{\cal Z}^1-{\cal Z}_2\bar{\cal Z}^2, \psi^1]
 &=& 2[{\cal Z}_2\bar{\cal Z}^1, \psi^2], \nonumber \\
 \frac k{2\pi}\gamma^\mu{\cal D}_\mu\psi^2
 +[{\cal Z}_1\bar{\cal Z}^1-{\cal Z}_2\bar{\cal Z}^2, \psi^2]
 &=& 0, \nonumber \\
 \frac k{2\pi}\gamma^\mu{\cal D}_\mu\bar\psi_1
 +[{\cal Z}_1\bar{\cal Z}^1-{\cal Z}_2\bar{\cal Z}^2, \bar\psi_1]
 &=& 0, \nonumber \\
 \frac k{2\pi}\gamma^\mu{\cal D}_\mu\bar\psi^2
 -[{\cal Z}_1\bar{\cal Z}^1-{\cal Z}_2\bar{\cal Z}^2, \bar\psi^2]
 &=& -2[\bar{\cal Z}^1{\cal Z}_2, \bar\psi_1].
\end{eqnarray}
The second and the third equations are (\ref{psi0}) with the
second term sign flipped, so they can only be solved by
$\psi^2=\bar\psi_1=0$. Inserting $(\psi^1,\psi^2)=(\psi,0)$ or
$(\bar\psi_1,\bar\psi_2)=(0,\psi)$ to the first or the fourth
equations we get (\ref{psi0}).

Summarizing, for each monopole instanton in generic vacuum,
there are eight fermion zero-modes, with two each from
\[
 \Psi^1,~~~\bar\Psi_2,~~~Z_1\bar\Psi_3-\bar Z^2\Psi^4,~~~
 Z_1\bar\Psi_4+\bar Z^2\Psi^3.
\]
This  clearly reduces to the previous result
for monopole instantons when $Z_2=0$.
In this generic vacuum, the zero-modes of a single monopole
are in one-to-one correspondence with the eight broken
supercharges. Although the number of the broken
supersymmetry is only six for monopoles in the special
vacua, the number of zero-modes cannot change just by
choice of the vacuum.
The eight zero-modes per monopole therefore persist
in all broken vacua, generic or special. This explains
why we found eight zero-modes in the previous section,
despite the half-BPS nature.

\section{The Vertex Operator and Non-perturbative Effective \\ Action}

The monopole instanton will contribute
a local operator to the effective action.
The purpose of this section is to discuss the possible
form of such non-perturbative terms in the effective action.
However, with the Chern-Simons term present,
there is a subtlety one must first understand.

There is a well-known argument \cite{Affleck:1989qf} that seemingly forbid
the monopole instanton contribution to the Euclidean path integral
for generic Chern-Simons theory.
As a simple example, let us recall once again the $SU(2)$
Chern-Simons theory with an adjoint scalar $\Phi$.
We discussed in Section 4 how the Chern-Simons action transforms
under complexified gauge transformations.
Let us consider here the real gauge transformation of the form
\begin{equation}
 g=e^{i\lambda\Phi}.
\label{g-lambda}
\end{equation}
The scalar field is invariant under this, while the Euclidean
action for $m$-monopole background is shifted by a pure imaginary constant,
\begin{equation}
\delta S_{\rm CS}=ikm\lambda,
\end{equation}
where we used ${\rm tr}\vev\Phi^2=\frac12$.
Now for $\lambda\; \slash \hskip-4.5mm\in 2\pi \mathbb Z$,
$\lambda$ is neither
a small gauge transformation nor a large gauge transformation,
so the path-integral over all gauge field configuration implies
integral over the gauge orbit, in other words an integral
over $\lambda$ from 0 to $2\pi$.
This seemingly projects out the contributions from the
sectors with nonzero monopole number.

This argument, however, overlooks another
important aspect of the Chern-Simons theory,
where the Gauss constraint relates flux to electric charge.
A monopole instanton induces a jump in total magnetic flux,
and must be accompanied by a related jump in total electric charge.
The final state and the initial state,
mediated by the monopole instanton, differ
by an $U(1)$ electric charge $\sim km$.
The constant gauge transformation by $\lambda(\infty)$ measures
precisely this electric charge, so the product of wavefunctions
also transform by a phase $e^{-ikm\lambda(\infty)}$.
The transition amplitudes for monopole-mediated processes are
therefore {\it not} projected out by integrating over the gauge orbit.

One can show the full gauge invariance of monopole-mediated amplitudes
by taking account of the gauge variance of the Lagrangian carefully.
To understand how to proceed, let us regard the system as a mechanical
system with a dynamical variable $q(t)$ and the Lagrangian $L[q]$.
Suppose the equation of motion is invariant under a group of
symmetry transformations $G$, but $L$ is invariant only up to total
time derivative.
\begin{equation}
 g\in G~:~
 q\longmapsto q^g\,,~~~
 L[q]\longmapsto L[q^g]=L[q]-\frac{d}{dt}2\pi\alpha_1[q,g]\,.
\label{1coc}
\end{equation}
The functional $\alpha_1$ is called {\it 1-cocycle} due to
the composition rule,
\begin{equation}
 \alpha_1[q,g]+\alpha_1[q^g,g']~=~ \alpha_1[q,gg'].
\end{equation}
The Noether charge gets modified due to this last term of (\ref{1coc}),
so that the corresponding quantum operator $g$ acts
on the basis states as follows,
\begin{equation}
 \bra{q}g= e^{2\pi i\alpha_1(q,g)}\bra{q^g}\,.
\label{qg}
\end{equation}
Now consider the transition amplitude between the states
$\ket{\Psi_i}$ and $\bra{\Psi_f}$ whose wave packets are localized
near $q=q_i$ and $q=q_f$.
The path integral gives
\begin{equation}
 \vev{\Psi_f|e^{-iH(t_f-t_i)}|\Psi_i}
~=~
 \int [dq]
 \Psi_f^\ast[q(t_f)]\Psi_i[q(t_i)]\exp\big(iS(t_f;t_i)\big)\,.
\end{equation}
One can compute the kernel $\bra{q_f}e^{-iH(t_f-t_i)}\ket{q_i}$
approximately using the classical action for a stationary path
connecting $q(t_i)=q_i$ and $q(t_f)=q_f$.
The kernel is not invariant under $G$ due to (\ref{1coc}).
Also, $G$-transformation of wave functions gives rise to
a phase factor due to (\ref{qg}):
the wave functions for the states $\ket{\Psi}$
and $\ket{\Psi^g}\equiv g\ket{\Psi}$ are related via
\begin{equation}
 \Psi^g(q)=\Psi(q^g)e^{2\pi i\alpha_1(q,g)}\,.
\end{equation}
The phase rotations of the kernel and wave functions cancel, so
that the transition amplitudes are invariant.
When applied to the previous Chern-Simons theory example and
$g$ is chosen to be a constant gauge transformation,
these phase rotations reflect the flux of monopole instanton
and the charges of the states.

The 1-cocycle $\alpha_1$ is trivial if it is solved in terms a
{\it 0-cocycle} functional $\alpha_0$,
\begin{equation}
 \alpha_1[q,g]=\alpha_0[q^g]-\alpha_0[q],
\label{0coc}
\end{equation}
since the theory is then described by a fully $G$-invariant
Lagrangian
\begin{equation}
\tilde L[q]~=~ L[q]+\frac d{dt}2\pi\alpha_0[q]\,,
\end{equation}
and the wave functions $\tilde\Psi(q)=\Psi(q)e^{2\pi i\alpha_0[q]}$
satisfying $\tilde\Psi^g(q)=\tilde\Psi(q^g)$.
However, in Chern-Simons theories the 1-cocycle can only formally
be solved, and the resulting 0-cocycle turns out to be
a nonlocal functional \cite{Dunne:1989cz,Elitzur:1989nr}.
In Appendix A we record an explicit form of $\alpha_0$ for
a simple BF-matter theory.

Let us turn to discuss in some detail the gauge transformation property
of our monopole solution in the $U(2)\times U(2)$ ABJM model.
In Sections 3 and 5 we solved the equations of motion
by a simple embedding of the 't Hooft Polyakov monopole $({\cal A},\Phi)$.
The embedding is such that the classical Chern-Simons action for the
two $U(2)$ gauge fields cancel, but the scalars ${\cal Z}_\alpha$ and
$\bar{\cal Z}^\alpha$ are not conjugate of each other at infinity.
A complexified gauge transformation can correct this wrong
asymptotics, but makes the total Chern-Simons action non-vanishing.
The end result was $e^{-S_{\rm E}}=(u_1/u_2)^k$ for one monopole
where $u_1, u_2$ are the eigenvalues of $\vev{Z_1}$.
Speaking in terms of cocycles, what we have done is to use the
1-cocycle relation (\ref{1coc}) to relate the values of classical
action in a ``wrong gauge'' to a ``real gauge''.

The Euclidean action is therefore not invariant
under some gauge transformations.
Indeed, the vevs of $Z_\alpha$ are simultaneously diagonalizable
and in general break the gauge group from $U(2)\times U(2)$
down to $U(1)^4$.
A $U(1)^2$ subgroup rotates $u_1$ and $u_2$ by independent phases,
and shifts the Euclidean action by pure imaginary constant.
The monopoles carry charges under the $U(1)$ group which phase-rotates
$u_1$ and $u_2$ oppositely.

However, this does not imply the monopole effect is projected out,
because we are not integrating over this gauge orbit.
As was reviewed in Section 2, the moduli space of vacua is
$({\mathbb C}^4/{\mathbb Z}_k)^2/S_2$, and in particular
the two vacua labelled by $(u_1,u_2)$ and
$(e^{-i\lambda_1/k}u_1,e^{-i\lambda_2/k}u_2)$ are not
gauge equivalent unless $\lambda_i\in 2\pi{\mathbb Z}$.
This is precisely because of the monopole-instantons
breaking $U(1)^2$ down to $({\mathbb Z}_k)^2$.
Our monopole-instanton action is clearly invariant under this
orbifold group, and it can be lifted to a well-defined
function on the moduli space.

Thus we can find the instanton contribution to the
effective action, weighted by $e^{-S_{\rm E}}$. As emphasized before, the
monopole-instanton carries the electric charges in addition to the
creation or annihilation of certain magnetic flux. We therefore consider
the charge-flux creation operator, or simply vertex operator, to describe
the effective interactions induced by those instantons.
The charge creation operators are in general
non-local operators because of their long-range electric fields.
In the Chern-Simons theories, however, the electrically charged states
do not emit the electric field, but are tied with local magnetic flux.
It implies that the charge-flux creation operators can now become local. For
example, local gauge-invariant charge-flux creation operators
for scalar fields $z_\a^i$ are given by $\tilde z_\a^i$
(\ref{correctscalar}),
\begin{eqnarray}\label{gaugeinscalar}
  \tilde z_\a^1 = e^{-i\th/k +i\s/2k} z_\a^1,
  \qquad \tilde z_\a^2 = e^{-i\th /k-i\s/2k} z_\a^2\ ,
\end{eqnarray}
where $\th=\frac12(\th^1+\th^2)$ and $\s=\th^2-\th^1$.
It is the $\s$ normalized to have period $2\pi$
that properly describes  the effect of the
monopole-instantons.
Some details are explained in Appendix C.

The vertex for the instanton has to do two things.
First it should create or destroy certain quantized magnetic flux,
which can be written in terms of a dual photon
field $\sigma$ as $e^{im\sigma}$.
In Appendix C, we show that this is indeed the case.
Thus the rough form of the gauge-invariant
vertex is ($m>0$)
\begin{equation}\label{monvertex}
e^{-S_{\rm E}+im\sigma } = \left(\frac{z^1}{z
^2}\right)^{km} e^{im\sigma} = \left( \frac{\tilde z^1}{\tilde z^2}
\right)^{km}
\end{equation}
since our notation is such that $\langle z^i\rangle = u_i$.
Second, the vertex must also carry $km$ units of an electric charge.
For $m>0$, the vertex we wrote already reflects this since
$z^1_\alpha$ and $z^2_\alpha$ are oppositely charged at unit $\pm 1/2$.

Incorporating the effect of fermionic zero-modes and the
conformal invariance, we expect further prefactors from
zero-modes and massive modes. The net effect is to have
additional nonperturbative corrections to the effective
Lagrangian in the broken phase,
\begin{equation}
{\cal L}_{\text{non-perturbative}}=\sum_m \left({g_{k,m}(
\tilde z,\tilde{ \bar z},\nabla \tilde z,\nabla \tilde{\bar z},
\tilde \Psi, \tilde{\bar \Psi})}
\left(\frac{z_1}{z_2}\right)^{km} e^{im\sigma}+c.c\right)
\end{equation}
where $g_{k,m}$ are dimension-three and charge-neutral operators.
When we do not consider motion of the vacuum moduli ($\nabla \tilde z=0
=\nabla \tilde{\bar z}$),
the only possible term is the eight-fermion term with
\begin{equation}
  g_{k,m}\sim
\frac{f_8(\tilde\Psi,\tilde{\bar\Psi},\tilde z,\tilde{\bar z})}
     {\mu^5(\tilde z,\tilde{\bar z})} \ ,
\end{equation}
where $\mu(\tilde z, \tilde{\bar z})= \mu( z, {\bar z})$ denotes the unique mass
parameter (\ref{mass0}) on the vacuum moduli space and $f_8$ is an
8-th order polynomial in the fermions with dependence on the
scalar vev only through ratios. The charge-neutrality here implies
that $f_8(\tilde\Psi,\tilde{\bar\Psi},\tilde z,\tilde{\bar z})
=f_8(\Psi,{\bar\Psi},z,{\bar z})$.
Here we indicated only the
rough scaling behavior. One can further restrict the possible
structure of this term from the non-anomalous $SU(4)_{\rm R}$
symmetry.

If we allow motion of the vacuum moduli,
we will have various mixing terms
between fermions and $\nabla z$, $\nabla\bar z$.
Recalling the discussion in \cite{Polchinski:1997pz}
about eight fermion zero-modes,
we believe that the purely bosonic terms generated by instanton effects
should start with four-derivatives
\begin{equation} \label{four}
g_{k,m}\sim\frac{|\nabla \tilde z|^4}{\mu^3(\tilde z,\tilde{\bar z})}
 \end{equation}
again up to a dimensionless neutral operator.
Determining the structure of these vertex operators
in full detail is beyond the scope of this work
and needs more careful analysis. In the next section, we will try to
compare the four-derivative terms (\ref{four})
to those in the dual supergravity picture.

\section{M/IIA Bulk Computation}

The $U(N)\times U(N)$ ABJM model is believed to be the
worldvolume theory of $N$ M2-branes in $\IC^4/\IZ_k$
orbifold. This proposal is so far supported by several basic
evidences. One is that the theory has the right supersymmetry
and conformal symmetry. Another is that the massless degrees of
freedom of the ABJM model match precisely with those of the
nonlinear sigma models of such M2-branes. Also it has been shown
that counting of superconformal indices
\cite{Bhattacharya:2008bja} is consistent with this proposal.

However, there is also a potentially contradicting piece
of evidence that seems to say that the number of isolated vacua of
a mass-deformed ABJM model is different from what is
expected from the bulk side in the large $N$ limit \cite{Gomis:2008vc}.
Given these mixed results, it is natural to ask whether we
can find further supporting evidences by considering
more sophisticated aspects of the theory, such as quantum-corrected
interactions.

An interesting analog can be found by considering D2-branes
on flat $\mathbb{R}^7$, which are nothing but M2-branes
on $S^1\times \IR^7$. The worldvolume theory of multiple D2-branes
is given by ${\cal N}=8$ $U(N)$ Yang-Mills theory, where monopole
instantons of usual kind exist in the Coulomb phase where
an adjoint scalar $\phi$ takes a vev.
Polchinski and Pouliot \cite{Polchinski:1997pz} computed,
for the case of $U(2)$,
what kind of interactions are generated by these instantons
and found four-derivative terms, such as
$e^{-4\pi\phi/ e^2}(\nabla\phi)^4/\phi^3$,
and its supermultiplet up to eight fermion terms, suppressed
exponentially by the Euclidean action of the instanton.

On the other hand, since D2-branes are really M2-branes,
a pair of D2-branes separated by a
distance $r$ in the IIA theory exchange 11-dimensional
supergravitons. In particular, when the
momenta being exchanged are those associated with the
 M-theory circle $S^1$, this generates quantum
correction of type $e^{-mr/R}(\nabla r)^4/r^3$
where $R$ is the radius of the eleventh circle
and $m$ is a positive integer.
Alternatively we can think of this process as exchange
of $m$ D0-branes.
Since we can interpret the worldvolume quantities as
$\alpha'\phi\sim r$ and  $e^2\alpha' \sim R$,
this interaction term computed from M-theory is exactly
the same as the four-derivative monopole instanton vertex
above computed from ${\cal N}=8$ Yang-Mills theory.

Here we would like to make a similar comparison for the ABJM
proposal of M2-branes. In previous sections, we already discussed
how the monopole instantons lead to quantum correction to the
effective Lagrangian at the level of four-derivative terms
and the supermultiplet thereof. Although we did not derive the
exact form of the vertex, we did derive the leading $k$-dependence
of the vertex and also how it scales with the mass scale $\mu$
of the generic Coulombic vacua. In the following, we will compare
these four-derivative vertices to those found in the bulk
computation where M2-branes scatter off each other via M-theory
supergraviton exchange or alternatively where D2-branes interact
via exchange of D0-branes.

\subsection{M-Theory Picture: Four-Derivative Interactions}

\begin{figure}[t]
\begin{center}
\includegraphics[width=12cm]{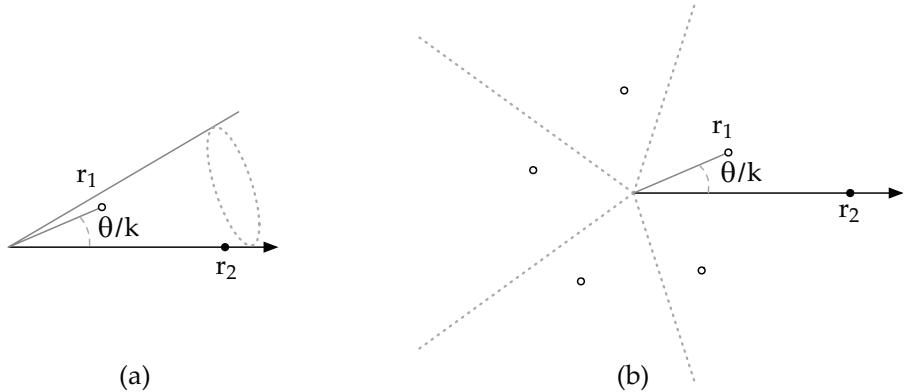}
\caption{(a) Two M2-branes placed in the $\IC/\IZ_k$ subspace
of the cone $\IC^4/\IZ_k$. (b) The covering space view of the same
configuration. } \label{cone1}
\end{center}
\end{figure}

We think of the two M2-branes as a source and a probe. The source
produces a background field configuration,
\be
\label{source-met}
ds^2 = h^{-2/3}dx^2_{1+2} + h^{1/3} dy^2 \,,
\;\;\;\;\;
C_{012} = h^{-1} \,,
\ee
where the harmonic function for a single M2-brane is given by
\be
h = 1+ \frac{32\pi^2}{(M_{11}r)^6} \,.
\ee
Before proceeding further, however, we wish to argue that
the right thing to do to make a comparison against the gauge
theory result is to drop ``$1$'' in the harmonic function.

One way to achieve this naturally is to consider the number of
``source'' M2-branes to be very large and take the near horizon limit.
On the field theory side, this amounts to considering $U(N+1)$
theory broken to $U(N)\times U(1)$. The latter would involve
further complication due to the fact that the monopole instanton
carries $U(N)$ charge, which we would like to avoid.

Another is to compute everything as it is and then extrapolate
to small $r$ regime, while maintaining the velocity of M2-branes
also sufficiently small.
A priori, there is no overlap between the regime where this
bulk computation is trustable (i.e., long distance regime) and the regime
where the worldvolume gauge theory computation is reliable
(i.e., short distance regime). Nevertheless, with enough
supersymmetry, the structure of interactions mediated by
BPS objects tends to be preserved across such interpolations.
This has been seen time and again in the development of
D-brane physics. We will be testing the ABJM
proposal against the bulk computation, in this sense. Performing
such an extrapolation carefully is equivalent to using
\be
\label{h1x}
h = \frac{32\pi^2}{(M_{11}r)^6} \,
\ee
from the very start.

We then study the dynamics of the probe brane using
\be
S_{\rm probe} = T_2 \int d^3x \left( -\sqrt{-g} + \frac{1}{6} \e^{\m\n\l}
\p_\m X^M \p_\n X^N \p_\l X^P C_{MNP} \right) \,,
\ee
where $g_{\m\n}$ here is the pull-back of the metric (\ref{source-met}).
We will focus on a slow motion in a direction
transverse to the M2-brane worldvolume and perpendicular
to the separation between the two branes,
following a similar computation in flat background
\cite{Hyun:1998qf,Hyun:1999hm}.

{}From this, we anticipate to reproduce four-derivative terms
such as in Eq.~(\ref{four}). Given the 2+1-dimensional Lorentz
invariance, however, it suffices to consider uniform motion of
the M2-branes, encoded in velocities $v=\partial_t X$, instead
of considering $\nabla X$.
Expanding the probe action in powers of velocity $v$,
we find that the $(v^0)$ term vanishes due to the BPS cancellation
between the two terms. The $(v^2)$ term serves as the kinetic term
and the $(v^4)$ term is the leading interaction term. Explicitly,
the action up to the $(v^4)$ term is given by
\footnote{
The intermediate step goes like :$
\;\;\;-h^{-1}\sqrt{1-hv^2}+h^{-1} = \half v^2 + \frac{1}{8} h v^4 + \CO(v^6).
$
}
\be
\label{probe-v4}
S_{\rm probe} =  \int d^3 x \left[ \half T_{2} v^2 + \frac{1}{8} T_2 h v^4
+\CO(v^6) \right].
\ee

Suppose the two M2-branes are located at $\vec{z}$ and $\vec{w}$ in $\IC^4$.
Without loss of generality, we may assume $|\vec{w}|>|\vec{z}|$, and define
\be
x\equiv \frac{|\vec{z}|}{|\vec{w}|} < 1,
\;\;\;\;\;
y e^{i\sigma/k} \equiv \frac{\vec{z}^* \cdot \vec{w}}{|\vec{z}||\vec{w}|}
\;\;\; (0\le y \le 1, 0 \le \sigma \le 2\pi k) \,.
\ee
For a later comparison with  the field theory computation, it is
convenient to use the rescaled field theory variables
\be
\label{rescale}
Z_a^{\rm F.T.} = \sqrt{\frac{T_2}{2}} \left( X_{2a-1}+iX_{2a} \right)^{\rm Grav} =
\frac{M_{11}^{3/2}}{2\sqrt{2}\pi} \left( X_{2a-1}+iX_{2a} \right)^{\rm Grav}.
\;\;\; (a=1,\cdots, 4)
\ee
{}From now on we mean by $z$ and $w$ these rescaled coordinates of dimension 1/2.
The velocity $v$ is rescaled by the same factor to become a variable
of dimension $3/2$.

The $\IZ_k$ orbifolding introduces mirror images of $\vec{z}$ at
$e^{2\pi i \ell/k} \vec{z}$ $(\ell=1, \cdots k)$, so instead of having
a single $hv^4$ term, we will have $k$ copies of $h$ with rotated centers
contributing. This effectively replaces $h$ by (up to an overall normalization),
\be
F_{k}(\vec{z},\vec{w}) &\equiv&\sum_{\ell=1}^{k} |\vec w-e^{2\pi il/k}\vec z|^{-6}
\nonumber\\
&=& \sum_{\ell=1}^{k}
\frac{1}{\left(|\vec{z}|^2+|\vec{w}|^2-2|\vec{z}^*\cdot \vec{w}|
\cos(2\pi \ell/k + \sigma/k) \right)^3} \,,
\ee
which reduces the
periodicity of the harmonic function to $2\pi$. The angular
coordinate $\sigma$ is to be identified with the dual photon field
that makes appearance in the monopole instanton vertex, and the
$m$-instanton amplitude is expected to be proportional to the
$m$-th Fourier coefficients of $F_{k}(\vec{z},\vec{w})$; \be
F_{k}(\sigma) = \sum_{m=-\infty}^{\infty} f_{k,m} e^{i m\sigma}\ .
\ee Each and every summand represents the monopole vertex of type
(\ref{four}). This is an expansion of the four-derivative
interaction between a pair of the M2-branes, in terms of the
angular momentum $m$ of the angle $\sigma$. In type IIA
interpretation, as we will see later, $m$ labels the number of
D0-branes being exchanged by the pair of D2-branes. D0-brane is
still the Kaluza-Klein momentum of the 11-th direction, although
the latter is now an azimuthal angle rather than a topological
circle. Collecting the results, we find the following effective
action in terms of the field theory variables
\be
S_{\rm probe} = \int d^3 x \left[ v^2 + \frac{ v^4}{8\pi^2}
\Big( f_{k,0}(\vec{z},\vec{w}) + \sum_{m=1}^{\infty} f_{k,m}(\vec{z},\vec{w}) (e^{im\sigma}+e^{-im\sigma})\Big)
  \right] \, ,
\ee
up to order $v^6$.

We thus find the M-theory counterpart of
(\ref{four}) as
\be
 \frac{v^4}{8\pi^2} f_{k,m} e^{i m\sigma}\ ,~~~~~
f_{k,m} = \int_0^{2\pi} \frac{d\sigma'}{2\pi} \,F_{k}(\sigma') e^{-i m\sigma'}
\,,
\ee
where the overall normalization is fixed by combining
(\ref{h1x}) and (\ref{probe-v4}) and taking the rescaling (\ref{rescale})
into account.
We can combine the $\sigma'$-integral and the sum over mirror images
into an integral over the circle in the ``covering space'' ($\sigma'/k\goto \beta$),
\be\label{partialwave}
f_{k,m} (\vec{z},\vec{w}) = k \int_0^{2\pi} \frac{d\beta}{2\pi}
\frac{e^{-imk\beta}}{\left(|\vec{z}|^2+|\vec{w}|^2 -2|\vec{z}^*\cdot \vec{w}| \cos\beta\right)^3}\, .
\ee
The integral can be most easily evaluated by a contour integral
along a unit circle on the complex plane ($e^{i\beta} \goto z$).
The result is
\be
\label{hsum}
 f_{k,m}(\vec{z},\vec{w})  =
\frac{ 8\pi^2q^{k|m|} }{(2\pi/k)^3(q^{-1}-q)^3|\vec{z}^*\cdot\vec{w}|^3} \cdot a_{k,m}(q)\,,
\ee
where $q<1$ is defined by
\be
\label{qxy}
q+ \frac{1}{q} = \frac{1}{y} \left(x +\frac{1}{x} \right) \, ,
\ee
and
\be\label{coeff}
a_{k,m}(q)=
\frac{\pi m^2}{2} + \frac{3\pi |m| (1+q^2) }{2|k| (1-q^2)}
+\frac{2\pi(1+4q^2+q^4)}{k^2 (1-q^2)^2}  \,.
\ee

The match with the field theory counterpart is easily seen
by noting that the parametrization of the generic vacuum
(\ref{generic-vev}) translates to
\be
\label{vev-grav}
\vec{z} = (u_1, cu_2,0,0),
\;\;\;
\vec{w} = (u_2, cu_1,0,0).
\ee
With this choice, the relation (\ref{qxy}) yields $q= u_1/u_2$.
So one can identify the suppression factor (exponential in $k$)
as the Euclidean action
\be
e^{-S_{\rm E}} = q^{k|m|} \,,
\ee
which matches precisely with  the field theory analysis (\ref{SEft}).
Furthermore, the dependence on the fundamental scale is also
reproduced correctly since
\be
(2\pi/k)^3(q^{-1}-q)^3|\vec{z}^*\cdot\vec{w}|^3 =
\left(\frac{2\pi}{k}(1+c^2)(u_2^2-u_1^2)\right)^3=\mu^3\,.
\ee
Interestingly, the dependence on the variable $c$ appears
only through this mass scale term in (\ref{hsum}).  Thus the
transfer of $m$ unit of momenta along $\sigma$ direction
generates the following term in the probe M2-brane dynamics
\be
\label{hsum3}
\frac{v^4}{8\pi^2}f_{k,m}(\vec{z},\vec{w})e^{im\sigma} =
\frac{v^4q^{k|m|}e^{im\sigma}}{\mu^3}\,   a_{k,m}(q)\, .
\ee
This is consistent with the monopole instanton vertex in
Eq.~(\ref{four}). Thus,
we find that the ABJM field theory at the nonperturbative level
captures the behavior of multiple M2 brane physics faithfully.

The field theoretical computation can be further improved.
For instance, the above M-theory computation provides the exact
expression for the prefactor in the form of $a_{k,m}(q)$
in (\ref{coeff}), which captures the complicated dependence on
ratios of the vev. This, together with $1/\mu^3$ factor,
should match the higher order  corrections to the saddle point
approximation in the field theory side. Also $m=0$ term
in the effective Lagrangian, corresponding to the supergraviton
exchange in the sector where $\sigma$ momentum
is zero, should come from ordinary perturbative
corrections in the field theory side. More precise check
of the ABJM proposal should be possible by computing these two
classes of  quantum corrections.

\subsection{Consistency Check with IIA Picture}

\noindent{\it D0-brane probe in $\IC^4/\IZ_k$}

The bulk picture can be thought of in two equivalent ways.
In the above M-theory picture, we have $N$ M2-branes
in the orbifold $\IC^4/\IZ_k$. In the second, related picture, we have
$N$ D2-branes in $\IC\IP^3$ with nonconstant 11-th radius and
a nontrivial RR field strength $dC_1$. In the latter,
the series of interactions we found above can be understood as
exchange of different number of D0-branes between a pair of D2-branes.
Here, we will work in this latter picture and work out the
$k$ and $m$ dependence of the amplitude according to the
D0-brane exchange picture.

Since we have no compact $S^1$, one might wonder what
D0-branes are from the M-theory perspective. Note that
the above expansion of the M-theory effective Lagrangian
to sectors with different $m$ is nothing but expansion
of the full 11-dimensional amplitude into some
angular-momentum eigensectors. If we choose to label
the associated angle as the 11-th direction, the quanta of
its conjugate momentum should be called D0-branes. Even though this
 11-th direction does not define a topological circle,
it is still a Killing direction so that we have a conserved
conjugate momentum. IIA picture will see these quanta as D0-branes,
Here we wish to confirm whether the individual amplitudes
are consistent with the interpretation in terms of the D0-brane
worldline viewpoint.

We work in the $\IC^4/\IZ_k$ orbifold which is the vacuum
moduli space of the ABJM model. The metric is given by
\be
ds_{\rm M}^2 = dx^2_{1+2} + dr^2 + r^2\left\{ d\O^2_{\IC\IP^3}+\frac{1}{k^2}(d\psi+k\,C)^2 \right\} \,.
\label{Mbac}
\ee
We rescaled the angle of the $S^1$ fiber such that $\psi$ has
period $2\pi$. The 1-form $C$ satisfies $dC=2J$ where
$J$ is the standard K\"ahler form of $\IC\IP^3$.
KK reduction along the $\psi$ direction gives the IIA background,
\be
ds_{\rm IIA}^2 &=& \left(\frac{r}{k}\right) \left[ dx^2_{1+2} + dr^2 + r^2 d\O^2_{\IC\IP^3} \right] \,,
\nn \\
e^{2\phi} &=& \left(\frac{r}{k}\right)^3 , \;\;\;\;\;
C_{(1)} \;=\; k \,C \,.
\label{IIAbac}
\ee
Now, imagine placing two M2-branes in the geometry. We use the probe
approximation, that is, we neglect the back-reaction to the geometry.
Let $\vec{z}, \vec{w} \in \IC^4$ be the coordinates of the two M2-branes
in the covering space.

In the IIA picture, the instanton in question is a Euclidean D0-brane
connecting the two D2-branes.
The dynamics of the D0-brane should be captured by $e^{-S_{\rm DBI}+iS_{RR}}$, where
\be
S_{\rm DBI} &=& \int e^{-\phi} d \ell
\;=\; k \int \sqrt{(dr/r)^2 + ds^2_{\IC\IP^3}}\,,
\nonumber
\\
S_{\rm RR} &=& \int C_{(1)} = k \int C \,.
\label{D0-RR}
\ee
For simplicity, let us first focus on
the simple case where the two D2-branes are located
on the same point in $\IC\IP^3$, but separated in
the $r$-direction; take $c=0$ in equation~(\ref{vev-grav}).
Then, we find
\begin{equation}
e^{-|m|S_{\rm DBI}}=\left(\frac{u_1}{u_2}\right)^{k|m|} \,,
\end{equation}
which again coincides with the field theory result.

In general, with separation in $\IC\IP^3$, the problem gets more complicated
due to the presence of the RR-coupling (\ref{D0-RR}).
Let us sketch how a similar analysis goes through in this case.
We first notice that one can always move the two M2-branes to lie
in $\mathbb{C}^2\subset\mathbb{C}^4$ by using the $SU(4)$ rotation.
Using the standard coordinates,
\begin{equation}
 (z_1,z_2) = re^{i\psi}
 \Big(\cos(\theta/2)e^{i\phi/2},\sin(\theta/2)e^{-i\phi/2}\Big),
\label{polar}
\end{equation}
and dimensionally reducing along $\psi$ we get to the IIA picture.
The Euclidean D0-brane has the worldline action $S=ks$, where
\begin{equation}
 s ~=~ \frac12\int\left(
  \sqrt{dt^2+d\theta^2+\sin^2\theta d\phi^2} - i\cos\theta d\phi
 \right),~~~~\big(t\equiv\log(r^2)\big).
\end{equation}
The classical variational problem becomes well defined once we
Wick rotate the variable $\phi=i\varphi$ to make the action real.
The problem is to find the stationary path connecting two points
$(t_i,\theta_i,\varphi_i)$ and $(t_f,\theta_f,\varphi_f)$ with
$\varphi_i=\varphi_f=0$.
Using the explicit solution to the equation of motion, one can show
the classical action satisfies
\begin{equation}
 \cos((\theta_i-\theta_f)/2)\cosh s~=~\cosh((t_f-t_i)/2),
\end{equation}
which is in precise agreement with (\ref{qxy}).

\vskip10mm

\noindent{\it Mass of fundamental string}

Another important part of the four-derivative vertices is $1/\mu^3$
piece, which is determined by two considerations. First, $1/\mu^3$
carries the right dimension to render vertices to be of dimension three,
making the interaction conformal. Second, the massive
particles in the Coulombic vacua is set by the unique fundamental scale
$\mu$, so its appearance is natural. For the M2-brane interpretation
of the ABJM model to make sense, $\mu$ should correspond to the
mass of an open M2-brane wrapping M-theory circle and stretching
between the two M2-branes, or that of a fundamental string
stretched between the pair of D2-branes.
While this is an easy task, we show it here since we chose a rather
unconventional parameterization of the vev in the field theory.

The mass of a fundamental string stretched between two D2-branes
is given by\footnote{
Here we are using the standard polar coordinates for $\mathbb{R}^2$.
The coordinate $\theta$ here is different from that
in (\ref{polar}).
}
\begin{eqnarray}\label{energy}
  \mu_{bulk} = T_2 \left(\frac{2\pi}{k}\right) \int
  \sqrt{r^2 dr^2 + r^4 d \th^2}\,.
\end{eqnarray}
The curve that minimizes the mass is found to be
\begin{eqnarray}
  r^2(\th)=\frac{2a}{\sin{2(\th+\th_0)}}\ , \qquad
\end{eqnarray}
where $a$ and $\th_0$ are constants.
{}Using the boundary values (\ref{vev-grav}),
\begin{eqnarray}
  && (T_2/2) \,r^2(\th_1) = u_{1}^2 + c^2 u_{2}^2, \qquad \tan{\th_1} = \frac{c u_{2}}{u_{1}},
  \nonumber \\
  && (T_2/2) \, r^2(\th_2) = u_{2}^2 + c^2 u_{1}^2, \qquad \tan{\th_2} = \frac{c u_1}{u_2},
\end{eqnarray}
we can determine $a$ and $\th_0$,
\begin{eqnarray}
  T_2 \,a = 2 c u_1 u_2 , \qquad \th_0 = 0\ .
\end{eqnarray}
Inserting them back into the mass functional (\ref{energy}), we obtain
\begin{eqnarray}
  \mu_{bulk} &=& T_{2}  \left( \frac{2\pi}{k}\right) \int_{\th_2}^{\th_1} d\th \
  \frac{2a}{\sin^2{2\th}} \nonumber \\
  &=& T_{2}  \left( \frac{2\pi}{k}\right) \times a
  \left[ \frac{\cos{2\th_2}}{\sin{2\th_2}} -
  \frac{\cos{2\th_1}}{\sin{2\th_1}}\right] \nonumber \\
  &=&   \left( \frac{2\pi}{k}\right) (1+c^2)(u_2^2-u_1^2) \,,
\end{eqnarray}
in perfect agreement of the mass scale $\mu$ in the broken phase
of the field theory (\ref{mass0}).

\vskip 1cm
\centerline{\bf Acknowledgement} \vskip 5mm \noindent
We thank Pei-Ming Ho, Seok Kim, Hyeonjoon Shin, and Erick Weinberg
for discussions.
K.M.L., J.P., P.Y. are  supported in part by the KOSEF SRC Program
  through CQUeST at Sogang University.
K.M.L. is also supported in part by the KRF National Scholar program.
Sm.L. is supported in part by the KOSEF Grant R01-2006-000-10965-0
  and the Korea Research Foundation Grant KRF-2007-331-C00073.
J.P. is also supported in part by KOSEF Grant R01-2008-000-20370-0
  and by  the Stanford Institute for Theoretical Physics.
K.H. thanks the organizer of Summer Institute 2008 at Fuji-Yoshida, Japan
  for hospitality during his stay.
Sm.L. thanks the string theory group at National Taiwan University
  for hospitality during his visit.
J.P. and P.Y. also acknowledge the hospitality of the Aspen Center for Physics.

\vskip 1cm

\centerline{\Large \bf Appendix}

\appendix

\section{  Cocycles in a BF Theory }

In this section we illustrate how the gauge variance of
the Lagrangian can be improved by adding the 0-cocycle,
and how to obtain it by solving the Gauss constraint.
As a simple example, we consider the abelian BF-matter theory
which arises in the low-energy effective theory of the ABJM model.

It is important that the Lagrangian for Chern-Simons theories
is first order in time derivative.
The spatial components of the gauge fields are therefore divided into
canonical coordinates and momenta by a choice of polarization,
whereas the time components are Lagrange multipliers for the Gauss constraint.
The cocycle then depends also on the polarization,
recalling that the first order Lagrangian $L=p\dot q-H(p,q)$
transform under the canonical transformation $(q,p)\to (p,-q)$ as
\begin{equation}
 L'-L~=~ -q\dot p-p\dot q ~=~ -\frac{d}{dt}{(pq)}\,.
\end{equation}

Let us consider the BF-matter theory with the Lagrangian
\begin{equation}
 {\cal L}
  = -|D_\mu z|^2
    +\frac{ k}{4\pi}\Big(
      b_0(\partial_1 c_2-\partial_2 c_1)
    + c_0(\partial_1 b_2-\partial_2 b_1)
    + b_2\dot c_1 +c_2\dot b_1 \Big).
\end{equation}
The canonical coordinates are $z,b_1,c_1$,
and the commutation relation in the temporal gauge reads
\begin{equation}
 [c_i({\bf x}), b_j({\bf y})]_\text{ET}
 ~=~ \frac{4\pi i}{k} \epsilon_{ij} \delta^2({\bf x}-{\bf y}).
\end{equation}
The physical  wave function $\Phi(z,b_1,c_1)$ satisfies
the Gauss constraints,
\begin{eqnarray}
 \Big(  i \frac{\delta}{\delta \theta({\bf x})}
       -i \partial_1\frac{\delta}{\delta b_1({\bf x})}
       -\frac{k}{4\pi} \partial_2 c_1({\bf x})
 \Big) \Phi &=& 0,
 \nonumber \\
 \Big( -i\partial_1\frac{\delta}{\delta c_1({\bf x})}
       -\frac{k}{4\pi} \partial_2b_1({\bf x})
 \Big) \Phi &=& 0,
\end{eqnarray}
where $\theta$ is the canonical conjugate of the gauge charge density.
The solution is
\begin{equation}
\Phi ~=~
 \exp\Big\{\frac{ik}{4\pi}\int d^2{\bf x}\,
 c_1({\bf x})\partial_1^{-1}\partial_2b_1({\bf x})\Big\}\,
 \tilde{\Phi}(z({\bf x})e^{-i\partial_1^{-1}b_1({\bf x})} ).
\end{equation}
The exponential part is identified as the cocycle,
\begin{equation}
 2\pi\alpha_0(b_1,c_1) =
-\frac{k}{4\pi} \int d^2{\bf x}
 c_1({\bf x})\partial_1^{-1}\partial_2 b_1({\bf x}).
\end{equation}
Under the local gauge transformations, the action
$S=\int_{t_i}^{t_f}dt {\cal L}$ is not invariant,
but can be made invariant by adding the boundary terms from cocyles,
\begin{equation}
 {\cal S}_{\rm inv} ~\equiv~
  \int_{t_i}^{t_f}dtd^2{\bf x}\;{\cal L}\;
 +2\pi\alpha_0(b_1,c_1,t_f)
 -2\pi\alpha_0(b_1,c_1,t_i).
\end{equation}
The remaining part of the wave function
$\tilde{\Phi}(z({\bf x})e^{-i\partial_1^{-1}b_1({\bf x})})$
is invariant under local gauge transformation.
For states with charge $n$, $\tilde{\Phi}$  is a homogeneous function of order $n$.
The monopole action could have contributions from both $S_{\rm inv}$ and $\tilde{\Phi}$,
as one can see in the Appendix B.

\section{Complex Action   and   Monopole Action}

To acquaint the complex action and its stationary path,
let us consider a simple mechanics model
with a rotational symmetry.
With the periodic coordinate $\theta\sim \theta+2\pi$,
 its Lagrangian and Hamiltonian are
$L=r^2\dot{\theta}^2/2$ and $H=p^2/2r^2$, respectively,
where $p$ is the conserved angular momentum.
We are interested in calculating the amplitude
\begin{equation}
W =\frac{ \langle \Psi_f |e^{-HT} |\Psi_i \rangle}{ \langle \Psi_f|\Psi_f
\rangle^\frac12 \langle \Psi_i|\Psi_i\rangle^{\frac12}  }
\end{equation}
between initial and final states of angular momentum $p_f,p_i$.
We choose the wave functions to be functions of coordinate so that
$\Psi_i \sim  e^{ip_i\theta} $.
The norm of the initial and final wave functions are not relevant.
One can express the above amplitude as a path integral
\begin{equation}
\int [ d\theta] \Psi(\theta_f)^*
e^{-S_{\rm E}} \Psi(\theta_i) = \int [dpd\theta] e^{-S_{\rm E}-S_{\rm b}}\ ,
\end{equation}
where the Euclidean action and the boundary contribution are given by
\begin{equation}
 S_{\rm E} = \int d\tau\left(-ip\dot\theta+\frac{p^2}{2r^2}\right)  \ ,~~~~
 S_{\rm b} = i(p_f\theta_f-p_i\theta_i)\ .
\end{equation}
It is easy to find the stationary path of the above path integral.
{}From the $p,\theta$ variations, we get $p=ir^2\dot{\theta}$,
$\dot{p}=0$ and $p(t_{f,i})=p_{f,i}$. Note that the boundary
variations of $\theta_{f,i}$ fix the initial and the final momenta.
The solution is that $p_f=p_i$ and $\theta = -ip_i\tau/r^2$ up to a
constant shift of $\tau$. The total action becomes
\begin{equation} S_{\rm E}+ S_{\rm b}
= +\frac{p_i^2 T}{2r^2}\ . \end{equation}
  This is exactly what we expect
from an energy eigenstate of $E=p_i^2/{2r^2}$.  As
 $S_{\rm E}= -p_i^2/2r^2$, the wave function contribution is crucial. Note
that the stationary path of angle has an imaginary direction. One
point is that the phase is purely imaginary at the stationary
point and so that $e^{i\theta}$ and  $e^{-i\theta}$ are not
complex conjugate to each other along the stationary path.

We are applying the similar idea for our monopole instantons.
The partition function $Z$ can be written as
\begin{equation}  W= \int [d\phi]
\Psi_f(z^i)^*e^{-S_{\rm E}}\Psi_i(z^i)\,. \end{equation}
The monopole instanton is interpolating two states whose charge
difference is $km$ and so the vacuum wave function on $S^2_\infty$
is
\begin{equation} \Psi_f(z^i)^* \Psi_i (z^i) \sim \Big( \frac{z_1}{z_2} \Big)^{n_1}
\Big(\frac{\bar{z}_1}{\bar{z}_2}\Big)^{n_2}\ .
\end{equation}
We consider here only spatially homogeneous mode of the fields.
This carries zero charge under $z^i\rightarrow e^{i\epsilon}z^i$
and carries $km$ charge under $z^1\rightarrow e^{i\lambda} z^1,
z^2\rightarrow e^{-i\lambda}z^2$ if
\begin{equation}  n_1-n_2=km\,. \end{equation}
In terms of the   phase $\theta_i$ of the $z^i$ fields, the wave function becomes
\begin{equation}
  \langle \Psi_f|\Psi_i\rangle \sim e^{ km(i\theta_1-i\theta_2)}\,.
 \end{equation}
The modulus of the wave function cancels and does not appear in
the partition function.
In the wave function, there  would be also cocycles and
additional part linear in $b_1$ as presented in the previous section.

Now we consider the stationary configuration of the  Euclidean path integral.
We use the monopole solution  ${\cal Z}, \bar{\cal Z}, {\cal
A}=\tilde{\cal A}$ as the  field configuration and calculate
the action. This illuminates the finer points of the wave function
and cocycles.   In this case, the Euclidean Chern-Simons action
also vanishes. The wave function at infinity is almost abelian and
the cocycle will be approximated by the previous appendix,
\begin{equation}
S_{\rm E} + 2\pi i\alpha_0(\phi) + ikm (-\partial_1^{-1} b_1 + \theta )\,.
\end{equation}
The cocycle contribution vanishes since
it is linear in $b=A-\tilde{A}$ and $b$ vanishes for the present
field configuration.
The only possible contribution should arise from the wave function.

For the solution ${\cal Z}=\text(z_1,z_2) $ and $\bar{\cal
Z}=\text (\bar{z}_1,\bar{z}_2)$, the asymptotic value of the
solution from equation (\ref{scalars0}) becomes
\begin{eqnarray}  && \langle z_1 \rangle = u_1 e^{i\theta_1} =\sqrt{u_1u_2},
 \ \langle z_2 \rangle =u_2e^{i\theta_2} =\sqrt{u_1 u_2}
\,, \nonumber \\
&&  \langle \bar{ z}_1\rangle = u_1 e^{-i\theta_1}
=\sqrt{\frac{u_1}{u_2}}, \ \langle \bar{z}_2 \rangle
=u_2e^{-i\theta_2} =\sqrt{\frac{u_2}{u_1}}\,.
\end{eqnarray}
Thus the asymptotic value of the phase becomes imaginary
\begin{equation}  e^{-i\theta_1} = e^{i\theta_2} = e^{\Lambda_*} =
\sqrt{\frac{u_1}{u_2}}\,. \end{equation}
For the BPS  solutions ${\cal Z},\bar{\cal Z}, {\cal A}=\bar{\cal
A}$ of $m$ monopoles, the matter action, the Chern-Simons term and
the cocycles all vanish except the phase term from the wave
function which is imaginary, or
\begin{equation}  e^{-S_{\rm E}}  = e^{ikm(\theta_1-\theta_2)}  =
\Big(\frac{u_1}{u_2}\Big)^{km}\,. \end{equation}

\section{Monopole Vertex Operator in the  ABJM Model}

As discussed in Section 4 and also in \cite{Lee:1991ge}, the vertex operators are widely used to
describe the low-energy effective interactions induced by
monopole-instanton solutions.
For the ABJM model,
the monopole instanton  vertex operators  carry both magnetic flux and electric
charge and  would be different from  those
in three-dimensional Maxwell theory.
 We discuss in this section the monopole vertex operators
in more details with emphasis on their physical origin.

Let us start with the flux creation operator $\Omega({\bf x})$
in three-dimensional Maxwell theory whose UV description is
the Georgi-Glashow model. It is well-known that an operator
$\Omega({\bf x})$ creating flux ${\cal B}$ at a point ${\bf x}$
takes the form as
\begin{eqnarray}\label{fluxop}
  \Omega({\bf x}) = \exp{\left(i \frac{\cal B}{4\pi} \s({\bf x})
  \right)}\ ,
\end{eqnarray}
where $\s$ denotes the dual photon
\begin{eqnarray}
  F_{\mu\nu} = \frac{1}{4\pi} \e_{\mu\nu\rho} \partial^\rho \s\ ,
  \qquad [\s({\bf x}),\partial_0 \s({\bf y})]_\text{ET} =   16 \pi^2 i
  \d({\bf x}-{\bf y}) \ .
\end{eqnarray}
Here $\s$ is normalized to have period $2\pi$. One can
show that $\Omega({\bf x})$ creates a flux ${\cal B} $ at ${\bf x}$
using the relation $\partial_0 \s= 4\pi  F_{12}$ together with canonical
equal-time commutation relation,
\begin{eqnarray}
   [F_{12}({\bf x}), \Omega({\bf y})] = \frac{1}{4\pi}
   [\partial_0 \s({\bf x}), \Omega({\bf y})]
   = {\cal B}  \d({\bf x}-{\bf y})\Omega({\bf y})\ .
\end{eqnarray}
For the monopole-instanton that creates the flux $4\pi m$,
the vertex operator becomes
\begin{eqnarray}
  \Omega_\text{monopole}({\bf x}) = \exp{(i m \s({\bf x}))}\ .
\end{eqnarray}

We now in turn consider the flux creation operator in the ABJM model whose
low-energy dynamics can be effectively described as the BF-theory (\ref{BF}).
It is not guaranteed that the flux creation operator
in the BF-theory takes the same form as the previous one.
We will show this is still
the case. Let us restrict
our attentions on a simple and illustrative BF-model
\begin{eqnarray}
  {\cal L} = - |D_\mu z|^2 + \frac{k}{4\pi} \e^{\mu\nu\rho}
  b_\mu \partial_\nu c_\rho
\label{L-BF}
\end{eqnarray}
with $D_\mu z=\partial_\mu z - i b_\mu z$.
The canonical commutation relation reads
\begin{eqnarray}\label{etc}
  [b_i({\bf x}),c_j({\bf y})]_\text{ET} = + i \frac{4\pi}{k} \e_{ij}
  \d({\bf x}-{\bf y})\ ,
\end{eqnarray}
once we choose the temporal gauge $b_0=c_0=0$. The Gauss laws become
\begin{eqnarray}\label{BFgauss}
  \frac{k}{4\pi} F^{(+)}_{12} - \rho_B = 0\,, \qquad F^{(-)}_{12}=0\ ,
\end{eqnarray}
where $\rho_B$ denote the gauge charge density and $F^{(+)}=dc$,
$F^{(-)}=db$.
They simply imply that we can identify the flux $F^{(+)}$ as
the asymptotic unbroken $U(1)$ field of the instanton which carries
the electric charges. For the vertex operator of instanton,
we therefore construct a certain operator $\Omega({\bf x})$
that creates flux $F^{(+)}$ and charges.

In order to find out $\Omega({\bf x})$ of our interest,
we first introduce the Lagrangian multiplier
\begin{eqnarray}\label{BFaction}
  {\cal L} = - |D_\mu z|^2 + \frac{1}{8\pi}\e^{\mu\nu\rho}
  \left(k b_\mu + \partial_\mu \s \right) F^{(+)}_{\nu\rho}\ .
\end{eqnarray}
The modified Lagrangian is invariant under the $U(1)$ gauge symmetry
\begin{eqnarray}
  b_\m\rightarrow b_\m+\partial_\m \l\ , \hspace{0.3cm}
  \s \rightarrow \s - k \l\ , \hspace{0.3cm}
  z \rightarrow e^{i\l} z\ .
\end{eqnarray}
Since, from (\ref{BFaction}), we can identify $\frac{1}{4\pi}F^{(+)}_{12}$
as the conjugate momentum of dual photon $\s$, the flux creation operator
can be described as
\begin{eqnarray}
  \Omega^0 ({\bf x}) =
  \exp{\left( i \frac{\cal B}{4\pi} \s({\bf x})\right)}\ .
\end{eqnarray}
It however transforms under the gauge symmetry:
\begin{eqnarray}
  \Omega^0({\bf x}) \ \rightarrow \
  \exp{\Big(-i\frac{k {\cal  B}}{4\pi}
  \l({\bf x})\Big)}\Omega^0({\bf x})\ .
\end{eqnarray}
We therefore conclude that, for gauge-invariance,
the flux creation operator also needs the creation
of ${k {\cal B}}/{4\pi}$ units of charges so as to satisfy
the Gauss law (\ref{BFgauss}).
The gauge-invariant charge-flux creation operator $\Omega({\bf x})$
thus takes the following form
\begin{eqnarray}
  \Omega({\bf x}) = \Omega^0({\bf x}) \cdot {\cal Q} ({\bf x})\ ,
\end{eqnarray}
where ${\cal Q}({\bf x})$ carries the charges ${{\cal B}k}/{4\pi}$
so that its local gauge transformation is opposite to
that of $\Omega^0({\bf x})$.
For the monopole-instanton that creates the flux $4\pi m$,
the vertex operator becomes
\begin{eqnarray}
  \Omega_\text{monopole}({\bf x}) = \exp{\Big( i m \s({\bf x})\Big)} {\cal Q}({\bf x})\
  ,
\end{eqnarray}
where the operator ${\cal Q}({\bf x})$ creates charge of $m k$.

These ideas can be applied to the ABJM model to explain the
charge-flux creation operators (\ref{gaugeinscalar}) and
the monopole vertex operators (\ref{monvertex}).


\newpage

\begin{thebibliography}{99}
\parskip 0.0cm



\bibitem{Klebanov:1996un}
  I. R.~Klebanov and A. A.~Tseytlin,
  ``{\it Entropy of near-extremal black p-branes},''
  Nucl.\ Phys.\  B {\bf 475}, 164 (1996)
  [arXiv:hep-th/9604089].





\bibitem{Bagger:2006sk}
  J. Bagger and N. Lambert,
  ``{\it Modeling multiple M2's},''
  Phys.\ Rev.\  D {\bf 75}, 045020 (2007)
  [arXiv:hep-th/0611108].

\bibitem{Bagger:2007jr}
  J. Bagger and N. Lambert,
  ``{\it Gauge symmetry and supersymmetry of multiple M2-branes},''
  Phys.\ Rev.\  D {\bf 77}, 065008 (2008)
  arXiv:0711.0955 [hep-th].

\bibitem{Bagger:2007vi}
  J. Bagger and N. Lambert,
  ``{\it Comments On multiple M2-branes},''
  JHEP {\bf 0802}, 105 (2008)
  arXiv:0712.3738 [hep-th].

\bibitem{Gustavsson:2007vu}
  A. Gustavsson,
  ``{\it Algebraic structures on parallel M2-branes},''
  arXiv:0709.1260 [hep-th].

\bibitem{Gustavsson:2008dy}
  A.~Gustavsson,
  ``{\it Selfdual strings and loop space Nahm equations},''
  JHEP {\bf 0804}, 083 (2008)
  arXiv:0802.3456 [hep-th].



\bibitem{Gaiotto:2008sd}
  D.~Gaiotto and E.~Witten,
  ``{\it Janus configurations,
  Chern-Simons couplings, and the theta-angle in ${\cal N}=4$
  super Yang-Mills theory},''
  arXiv:0804.2907 [hep-th].

\bibitem{Hosomichi:2008jd}
  K.~Hosomichi, K.M.~Lee, S.~Lee, S.~Lee and J.~Park,
  ``{\it ${\cal N}=4$ superconformal Chern-Simons theories
  with hyper and twisted hyper multiplets},''
  JHEP {\bf 0807}, 091 (2008)
  arXiv:0805.3662 [hep-th].

\bibitem{Hosomichi:2008jb}
  K.~Hosomichi, K.M.~Lee, S.~Lee, S.~Lee and J.~Park,
  ``{\it ${\cal N}=5,6$ superconformal Chern-Simons theories
  and M2-branes on orbifolds},''
  arXiv:0806.4977 [hep-th].

\bibitem{Aharony:2008ug}
  O.~Aharony, O.~Bergman, D. L.~Jafferis and J.~Maldacena,
  ``{\it ${\cal N}=6$ superconformal Chern-Simons-matter theories,
  M2-branes and their gravity duals},''
  arXiv:0806.1218 [hep-th].

\bibitem{Bagger:2008se}
  J.~Bagger and N.~Lambert,
  ``{\it Three-algebras and ${\cal N}=6$ Chern-Simons gauge theories},''
  arXiv:0807.0163 [hep-th].


\bibitem{Bandres:2008ry}
  M.~A.~Bandres, A.~E.~Lipstein and J.~H.~Schwarz,
  ``{\it Studies of the ABJM Theory in a Formulation with Manifest SU(4)
  R-Symmetry,}''
  arXiv:0807.0880 [hep-th].

\bibitem{Schnabl:2008wj}
  M.~Schnabl and Y.~Tachikawa,
  ``{\it Classification of ${\cal N}=6$ superconformal theories of ABJM type},''
  arXiv:0807.1102 [hep-th].



\bibitem{Bergshoeff:2008bh}
  E. A.~Bergshoeff, O.~Hohm, D.~Roest, H.~Samtleben and E.~Sezgin,
  ``{\it The superconformal gaugings in three dimensions},''
  arXiv:0807.2841 [hep-th].

\bibitem{Gomis:2008cv}
  J.~Gomis, A.~J.~Salim and F.~Passerini,
  ``{\it Matrix theory of type IIB plane wave from membranes},''
  JHEP {\bf 0808}, 002 (2008)
  [arXiv:0804.2186 [hep-th]].

\bibitem{Hosomichi:2008qk}
  K.~Hosomichi, K.~M.~Lee and S.~Lee,
  ``{\it Mass-deformed Bagger-Lambert theory and its BPS objects},''
  arXiv:0804.2519 [hep-th].

\bibitem{Gomis:2008vc}
  J.~Gomis, D.~Rodriguez-Gomez, M.~Van Raamsdonk and H.~Verlinde,
  ``{\it A massive study of M2-brane proposals},''
  arXiv:0807.1074 [hep-th].

\bibitem{Aharony:2008gk}
  O.~Aharony, O.~Bergman and D. L.~Jafferis, ``{\it Fractional M2-branes},''
  arXiv:0807.4924 [hep-th].




\bibitem{Bhattacharya:2008bja}
  J.~Bhattacharya and S.~Minwalla,
  ``Superconformal indices for ${\cal N}=6$ Chern-Simons theories,''
  arXiv:0806.3251 [hep-th].



\bibitem{Nishioka:2008gz}
  T.~Nishioka and T.~Takayanagi,
  ``{\it On type IIA Penrose limit and ${\cal N}=6$ Chern-Simons theories},''
  JHEP {\bf 0808}, 001 (2008)
  [arXiv:0806.3391 [hep-th]].

\bibitem{Gaiotto:2008cg}
  D.~Gaiotto, S.~Giombi and X.~Yin,
  ``{\it Spin chains in ${\cal N}=6$ superconformal Chern-Simons-matter theory},''
  arXiv:0806.4589 [hep-th].

\bibitem{Grignani:2008is}
  G.~Grignani, T.~Harmark and M.~Orselli,
  ``{\it The SU(2) x SU(2) sector in the string dual of N=6 superconformal
  Chern-Simons theory,}''
  arXiv:0806.4959 [hep-th].

\bibitem{Minahan:2008hf}
  J.~A.~Minahan and K.~Zarembo,
  ``{\it The Bethe ansatz for superconformal Chern-Simons},''
  arXiv:0806.3951 [hep-th].

\bibitem{Arutyunov:2008if}
  G.~Arutyunov and S.~Frolov,
  ``{\it Superstrings on AdS$_4\times \mathbb{CP}^3$ as a coset sigma-model},''
  arXiv:0806.4940 [hep-th].

\bibitem{Stefanski:2008ik}
  B.~Stefanski,
  ``{\it Green-Schwarz action for Type IIA strings on AdS$_4\times
  \mathbb{CP}^3$},''
  arXiv:0806.4948 [hep-th].

\bibitem{Gromov:2008bz}
  N.~Gromov and P.~Vieira,
  ``{\it The AdS4/CFT3 algebraic curve},''
  arXiv:0807.0437 [hep-th].

\bibitem{Gromov:2008qe}
  N.~Gromov and P.~Vieira,
  ``{\it The all loop AdS4/CFT3 Bethe ansatz},''
  arXiv:0807.0777 [hep-th].

\bibitem{Ahn:2008aa}
  C.~Ahn and R.~I.~Nepomechie,
  ``{\it ${\cal N}=6$ super Chern-Simons theory S-matrix and all-loop Bethe ansatz
  equations},''
  arXiv:0807.1924 [hep-th].

\bibitem{Bak:2008cp}
  D.~Bak and S.~J.~Rey,
  ``{\it Integrable spin chain in superconformal Chern-Simons theory},''
  arXiv:0807.2063 [hep-th].





\bibitem{Polchinski:1997pz}
  J.~Polchinski and P.~Pouliot,
  ``{\it Membrane scattering with M-momentum transfer},''
  Phys.\ Rev.\  D {\bf 56}, 6601 (1997)
  [arXiv:hep-th/9704029].


\bibitem{Paban:1998mp}
  S.~Paban, S.~Sethi and M.~Stern,
  Adv.\ Theor.\ Math.\ Phys.\  {\bf 3} (1999) 343
  [arXiv:hep-th/9808119].



\bibitem{Hyun:1998qf}
  S.~Hyun, Y.~Kiem and H.~Shin,
  ``{\it Effective action for membrane dynamics in DLCQ M theory on a two-torus},''
  Phys.\ Rev.\  D {\bf 59}, 021901 (1999)
  [arXiv:hep-th/9808183].

\bibitem{Hyun:1999hm}
  S.~Hyun, Y.~Kiem and H.~Shin,
  ``{\it Non-perturbative membrane spin-orbit couplings in M/IIA theory},''
  Nucl.\ Phys.\  B {\bf 551}, 685 (1999)
  [arXiv:hep-th/9901105].






\bibitem{Affleck:1989qf}
  I.~Affleck, J.~A.~Harvey, L.~Palla and G.W.~Semenoff,
  ``{\it The Chern-Simons term versus the monopole},''
  Nucl.\ Phys.\  B {\bf 328}, 575 (1989).

\bibitem{Lee:1991ge}
  K. M.~Lee,
  ``{\it Charge violation by instantons in Chern-Simons theories},''
  Nucl.\ Phys.\  B {\bf 373}, 735 (1992).
















\bibitem{Lambert:2008et}
  N.~Lambert and D.~Tong,
  ``{\it Membranes on an Orbifold},''
  Phys.\ Rev.\ Lett.\  {\bf 101} (2008) 041602
  [arXiv:0804.1114 [hep-th]].

\bibitem{Distler:2008mk}
  J.~Distler, S.~Mukhi, C.~Papageorgakis and M.~Van Raamsdonk,
  ``{\it M2-branes on M-folds},''
  JHEP {\bf 0805} (2008) 038
  [arXiv:0804.1256 [hep-th]].

\bibitem{Berenstein:2008dc}
 D.~Berenstein and D.~Trancanelli,
  ``{\it Three-dimensional ${\cal N}=6$ SCFT's and their membrane dynamics},''
  arXiv:0808.2503 [hep-th].




\bibitem{Weinberg:2006rq}
  E.~J.~Weinberg and P.~Yi,
  ``{\it Magnetic monopole dynamics, supersymmetry, and duality,}''
  Phys.\ Rept.\  {\bf 438} (2007) 65
  [arXiv:hep-th/0609055].

\bibitem{Callias:1977kg}
  C.~Callias,
  ``{\it Index theorems on open spaces},''
  Commun.\ Math.\ Phys.\  {\bf 62}, 213 (1978).

\bibitem{Weinberg:1979zt}
  E.~J.~Weinberg,
  ``{\it Fundamental monopoles and multi-monopole solutions
  for arbitrary simple gauge groups},''
  Nucl.\ Phys.\  B {\bf 167}, 500 (1980).





\bibitem{Dunne:1989cz}
  G.~V.~Dunne, R.~Jackiw and C.~A.~Trugenberger,
  ``{\it Chern-Simons theory in the Schr\"{o}dinger representation},''
  Annals Phys.\  {\bf 194} (1989) 197.

\bibitem{Elitzur:1989nr}
  S.~Elitzur, G. W.~Moore, A.~Schwimmer and N.~Seiberg,
  ``{\it Remarks on the canonical quantization of
  the Chern-Simons-Witten theory},''
  Nucl.\ Phys.\  B {\bf 326} (1989) 108.



\end{thebibliography}
\end{document}